\documentclass[12pt]{article}
\usepackage{latexsym}
\usepackage{amsmath}
\usepackage{amssymb}
\usepackage{bbm}
\usepackage{graphicx}
\usepackage{subfigure}
\usepackage{dcolumn}
\usepackage{bm}
\usepackage{cite}
\usepackage{booktabs}
\usepackage{youngtab}
\usepackage{marginnote}
\usepackage{mathtools}
\usepackage{mathrsfs}
\usepackage{makecell}
\usepackage{float}
\def\cI{\mathcal{I}}
\usepackage[usenames,dvipsnames]{color}
\usepackage[]{hyperref}

\usepackage{tikz}
\usepackage[height=8.85in,width=6.45in]{geometry}

\usetikzlibrary{arrows}
\usepackage{fancyhdr}
\usepackage{xcolor}
\newsavebox\MBox

\usepackage{setspace}
\usepackage{array}
\newcolumntype{P}[1]{>{\centering\arraybackslash}p{#1}}
\def\dburl{\url{www.rossealtman.com/tcy}}

\newcommand{\be}{\begin{equation}}
\newcommand{\ee}{\end{equation}}
\newcommand{\bea}{\begin{eqnarray}}
\newcommand{\eea}{\end{eqnarray}}

\def\IZ{\mathbb{Z}}

\def\cA{\mathcal{A}}

\def\cF{\mathcal{F}}
\def\cI{\mathcal{I}}

\def\cO{\mathcal{O}}

\def\l:{\mathopen{:}\,}
\def\r:{\,\mathclose{:}}

\usepackage{authblk}

\title{Applying machine learning to the \\  Calabi-Yau orientifolds with string vacua}
\author[1]{Xin Gao\footnote{xingao@scu.edu.cn}}
\author[2,3]{Hao Zou\footnote{hzou@bimsa.cn}}
\affil[1]{College of Physics, Sichuan University, Chengdu 610065, China} 
\affil[2]{Yau Mathematical Sciences Center, Tsinghua University, Beijing 100084, China}
\affil[3]{Beijing Institute of Mathematical Sciences and Applications, Beijing 101408, China}
\date{}

\begin{document}
\maketitle	

\begin{abstract}

We use  machine learning technique to search the polytope which can result in an orientifold Calabi-Yau hypersurface and  the  ``{\it naive Type IIB string vacua}".  We show that neural networks can be trained to give a high accuracy for classifying the orientifold property and vacua based on the newly generated orientifold Calabi-Yau database with $h^{1,1}(X) \leq 6$ \cite{Altman:2021pyc}.  This indicates the orientifold symmetry may already be encoded in the polytope structure.  
   In the end, we  try to use the trained neural networks model to go beyond the database and predict the orientifold signal of polytope for higher $h^{1,1}(X)$.

   \end{abstract}

\clearpage
\tableofcontents
\section{Introduction}

Orientifold Calabi-Yau threefolds represent a rich phenomenological starting  point for the construction of concrete string models for both particle physics and cosmology.    There are lots of important properties like possible proper divisor exchange involutions, the classification and counts of orientifold planes under the involution,  the non-trivial Hodge number splitting, which were systematically studied recently in \cite{Altman:2021pyc}.

The authors of  \cite{Altman:2021pyc} construct an orientifold Calabi-Yau threefold database  in a systematic way for $h^{1,1}(X) \leq 6$  (\dburl) by considering  non-trivial $\mathbb{Z}_{2}$ divisor exchange involutions. The orientifold Calabi-Yau  database is built on the threefolds  database \cite{Altman:2014bfa} constructed  from the Kreuzer-Skarke list  \cite{Kreuzer:2000xy, database},   and   the earlier classificaion of divisor exchange involutions  \cite{Gao:2013pra}.  In  \cite{Altman:2021pyc}, constructing orientifold Calabi-Yau  involves several technical procedures which we  summarized in section \ref{sec:orientifold}.  This procedure include determining the topology for each individual  divisor,  identifying and classifying the proper non-trivial involutions  for each unique  Calabi-Yau hypersurface.  Each of the proper involution will result in an new orientifold Calabi-Yau manifold  with non-trivial odd equivariant cohomology $h^{1,1}_{-}(X/\sigma^{*})\neq 0$. The authors clarified all possible fixed loci under the proper involution, i.e, the locations of $O3$, $O5$ and $O7$-planes in the Type IIB projection. It was found that under the proper involutions one ends up with a majority of $O3/O7$-planes systems, most of which further admit aa \lq\lq {\it naive Type IIB string vacua}"  by  checking the D3 tadpole cancellation condition.

As we can see, constructing orientifold Calabi-Yau manifolds depends nontrivially on the underlying manifold data and it presents an interesting challenge for machine learning \cite{Nielsen:2015nn, Ruehle:2020jrk}. Machine learning has been a good implement in theoretical physics research and leads to fruitful results during the last couple of years. With the help of machine learning people are able to deal with problems with more computational efficiency, especially the problems involving big data, for example, study the landscape of string flux vacua \cite{Cole:2018emh,Cole:2019enn,Krippendorf:2021uxu,Cole:2021nnt, He:2021nag,Ruehle:2017mzq,Halverson:2019tkf, He:2020mgx, Bena:2021wyr, He:2021eiu} as well as F-theory compactifications \cite{Carifio:2017bov,Wang:2018rkk,Bies:2020gvf}. This technique allows people to learn lots of quantities of Calabi-Yau manifolds, from its  toric building blocks like the polytope structure   \cite{Krefl:2017yox, Bao:2021ofk} and triangulations \cite{Altman:2018zlc,Demirtas:2020dbm}, to the calculation of Hodge numbers \cite{Bull:2018uow,He:2018jtw,He:2020lbz,Erbin:2020tks}, numerical metrics \cite{Anderson:2020hux,Jejjala:2020wcc,Douglas:2020hpv,Larfors:2021pbb} and line bundle cohomologies \cite{Klaewer:2018sfl,Brodie:2019dfx}.  Besides, machine learning has also been applied to study 
and find certain structures on Calabi-Yau for model building \cite{Krippendorf:2020gny,Deen:2020dlf, Otsuka:2020nsk, Ashmore:2021rlc,Abel:2021ddu,Abel:2021rrj,Constantin:2021for, Harvey:2021oue}.

The newly established orientifold Calabi-Yau database  \cite{Altman:2021pyc} and its corresponding polytopes are the ideal data for machine learning in several aspects. First, the  explicit formulas  to determine an orientifold Calabi-Yau  are not known and calculations rely on complicated and  computationally intense algorithms  \cite{Altman:2021pyc}.  It involves several computational algebraic geometry packages combined to work together \cite{sage, DGPS, Blumenhagen:2010pv, cohomCalg:Implementation, Kreuzer:2002uu}. So such topological properties are an interesting and challenging playground for machine learning. It is very interesting to see whether the machine learning can avoid these difficult calculations and is capable of learning this particular interesting property to get the desired polytope which can result in an orientifold Calabi-Yau and further the  \lq\lq {\it naive Type IIB string vacua}".  
 
Second, it was conjectured that the orientifold symmetry (or more precisely the involution symmetry) on the Calabi-Yau hypersurface is already encoded in the polytope structure. It is similar to the fact that when calculating the Hodge number of Calabi-Yau hypersurface $X$, one only need information of the reflexive polytope without desingularization \cite{Batyrev:1993}, or triangulations in another word. So one may wonder whether one can determine a polytope can result in an orientifold Calabi-Yau hypersurface and naive string vacua or not,  before a detail calculation. We expect by utilizing the power of the machine learning we may get closer to this question in this paper.

Third, it is very difficult to scan the  Kreuzer-Skarke database to find  the orientifold Calabi-Yau with higher $h^{1,1}$.  This difficulty is threefold.  One is due to the exponential increased size of  the Kreuzer-Skarke database  when $h^{1,1}$ goes higher  \cite{Kreuzer:2000xy, database}.  For example, in \cite{Altman:2021pyc} the authors considered 22974 favorable polytopes in total while only for $h^{1,1}(X) = 7$ itself, there are 50376 polytopes.  Moreover, when $h^{1,1}$ increases it is more difficult to get all triangulations of the polytope due to the exponentially increased possible ways of doing that. Finally, the number of possible involutions also increase exponentially and for some of them it is extremely slow to get the fixed loci. Putting all these difficulties together, it is very unlikely to scan all the Kreuzer-Skarke database in a brute force way to get the orientifold Calabi-Yau with an accessible computer power.  However, as  shown in \cite{Altman:2021pyc}, the percentage of polytope which can result in an orientifold Calabi-Yau  and  the \lq\lq {\it naive string vacua}" is very small (around 5\% for $h^{1,1} \leq 6$). Moreover this percentage  tends to decrease when $h^{1,1}$ goes higher. So the signal of orientifold is very rare in the full Kreuzer-Skarke database. This is exactly what we  want to try  to see whether the machine learning can help to pick out the \lq\lq orientifold" signal   in higher $h^{1,1}$.  Considering there is no concrete generic orientifold Calabi-Yau with high $h^{1,1}$, our efforts to explore such possibility using machine learning would be very helpful.  

 Although there are many benefits for the machine learning to do the prediction, one should note machine learning cannot solve the problem once and for all.   One has to check whether these prediction is correct or not. However, usually these check is hard to be done and the precision is not very high even the test accuracy is extremely high in training the neural network.  This is due to the fact one has to use a subset of the database to learn something more complicated,  just like in the Kreuzer-Skarke database,  the larger the $h^{1,1}$ is, the more complicated of the polytopes is. 

This paper is organized as follows. In section \ref{sec:orientifold}, we briefly summary the algorithm how to construct an orientifold Calabi-Yau manifold and the naive Type IIB string vacua. In section \ref{sec:ML} we apply the machine learning method to study the orientifold Calabi-Yau database  with $h^{1,1}(X) \leq 6$ \cite{Altman:2021pyc}. Since it gives a very high accuracy, we  will try to  apply our network model to explore the higher $h^{1,1}(X)$ case. In section \ref{sec: high} we do the first step to predict the polytopes in $h^{1,1}(X)=7$ which may give us the orientifold Calabi-Yau and vacua. We pick out some favorable cases to explicitly check whether our predictions give the right answer. We make a conclusion in section \ref{sec:con}.

\section{Construct the Orientifold Calabi-Yau}
\label{sec:orientifold}

Let us briefly recall some results from \cite{Altman:2021pyc} in which the standard procedure to identify an orientifold Calabi-Yau threefolds is described in detail. 

First we need a smooth description of our original Calabi-Yau hypersurface $X$ from the Kreuzer-Skarke database \cite{Kreuzer:2000xy, database}.  In doing so, we must at least partially desingularize the ambient toric variety, denoted as $\cA$, by blowing up enough of its singular points. A method for doing so is called maximal projective crepant partial (MPCP) desingularization, which involves the triangulation of the polar dual reflexive polytope $\Delta^{*}$,  containing at least one fine, star, regular triangulation (FSRT).  
We  define the MPCP-desingularized ambient 4D toric variety as:
\bea\label{eq:defamb}
\cA=\frac{\mathbb{C}^{k}\smallsetminus Z}{\left(\mathbb{C}^{*}\right)^{k-4}\times G}\, ,
\eea
where $Z$ is the locus of points in $\mathbb{C}^{k}$ ruled out by the Stanley-Reisner ideal $\mathcal{I}_{SR}(\cA)$, and $G$ is the stringy fundamental group (trivial in most cases,  there are only $14$ polytopes in $h^{1,1} \leq 6$ contain  non-trivial $G$). The geometry on this toric variety can be described by the projective coordinates $\{x_{1},...,x_{k}\}$ and their toric $\mathbb{C}^{*}$ equivalence classes
\bea
(x_{1},...,x_{k})\sim (\lambda^{\mathbf{W}_{i1}}x_{1},...,\lambda^{\mathbf{W}_{ik}}x_{k})\, ,
\eea
which define a projective weight matrix $\mathbf{W}$. However, there may  exist two or more  MPCP triangulations which result in the same Calabi-Yau hypersurface due to   Wall's theorem \cite{Wall}. This theorem shows the compact Calabi-Yau 3-folds are classified by their  Hodge numbers,  intersection numbers, and the second Chern Class.  This leads to a \lq\lq {\it Geometry-wise description}" in which  the various triangulations (phases of the complete K\"ahler cone) corresponding to a distinct Calabi-Yau threefold geometry were glued  together.
Furthermore, we restrict  ourselves to the so-called \lq\lq {\it favorable}" description, in which  the toric divisor classes on the Calabi-Yau hypersurface $X$ are all descended from ambient space $\cA$.

Starting from a favorable geometry-wise description, we need to identify the proper involution $\sigma$ which involves exchanging one or more pairs of divisors. Those divisors should have  the same topology   and at the same time have different weights (non-trivial identical divisors (NID)):
\bea
\sigma: x_i \leftrightarrow x_j \quad \Longrightarrow \quad \sigma^*: D_i \leftrightarrow D_j. \nonumber\\
H^{\bullet} (D_i) \cong H^{\bullet} (D_j),  \quad  \cO(D_i) \neq \cO(D_j)  
\eea
Furthermore, such involution should satisfy the symmetry of Stanley-Reisner Ideal  $\cI_{SR} (\cA)$ and the symmetry of   the linear ideal $\cI_{lin} (\cA)$. The first symmetry is  to ensure the involution be an automorphism of $\cA$, leaving invariant the exceptional divisors from resolved singularities. The later one ensures the defining polynomial of CY remains  homogeneous under involution. Putting these two together, the involution should be a symmetry of the Chow-group:
\bea
\label{eq:chow}
A^{\bullet}(\cA)  \cong \frac{\IZ ( D_1, \cdots, D_k)}{\mathcal{I}_{lin}(\cA)+ \mathcal{I}_{SR}(\cA)}\, , 
\eea
indicating the triple intersection form defined in the Chow-group is invariant under the involution $\sigma$. In this paper, we only consider the \lq\lq {\it geometry-wise proper involution}"  which are globally consistent across all disjoint phases of the K\"ahler cone for each unique Calabi-Yau geometry.  

The next task is to check whether there exist any point-wise fixed loci for a given involution on the Calabi-Yau threefold. The first step is to fix the invariant Calabi-Yau hypersurface polynomial  $P_{symm} = \sigma (P_{symm})$ and  the minimal generators $\mathcal{G}$  generated  by homogeneous polynomials  that are  (anti-)invariant under $\sigma$:
\bea
\mathcal{G}=\mathcal{G}_{0}\cup\mathcal{G}_{+}\cup\mathcal{G}_{-} \, .
\eea
The unexchanged coordinates in $\mathcal{G}_{0}$ are known  from our choice of involution. To find the non-trivial even and odd parity generators in $\mathcal{G}_{+}$ and $\mathcal{G}_{-}$,  we must consider not only $\sigma$, but all possible non-trivial ``subinvolutions'' $\rho\subseteq\sigma$ given by the nonempty subsets of $\{\sigma_{1},...,\sigma_{n}\}$ of size $1\leq m\leq n$. Then we denote the new coordinate in $\mathcal{G}\equiv\{y_{1},...,y_{k'}\} $ as:
$$
y_{\pm}(\mathbf{a})=  x_{i_{1}}^{a_{1}}x_{i_{2}}^{a_{2}}\dotsm x_{i_{m}}^{a_{m}} \pm x_{j_{1}}^{a_{1}}x_{j_{2}}^{a_{2}}\dotsm x_{j_{m}}^{a_{m}}\, ,
$$
The condition for { homogeneity}, in terms of the   columns $\mathbf{w}_{i_{s}}$ and $\mathbf{w}_{j_{s}}$ of the weight matrix $\mathbf{W}$ is given by:
\bea
a_{1}(\mathbf{w}_{i_{1}}-\mathbf{w}_{j_{1}}) + a_{2}(\mathbf{w}_{i_{2}}-\mathbf{w}_{j_{2}}) + \dotsb + a_{m}(\mathbf{w}_{i_{m}}-\mathbf{w}_{j_{m}})=0\, .
\eea

 The second step is to   perform a  Segre embedding transforming the projective coordinates into the (anti-)invariant generators $\{x_{1},...,x_{k}\}\mapsto\{y_{1},...,y_{k'}\}$ which constructs a new weight matrix $\tilde{\mathbf{W}}$ for $\{ y_i \}$.  Then we can find out the naive fixed point loci in the new weight matrix.  
In order for a codimension-1 subvariety $D\subset X$ to be point-wise fixed under the involution, the corresponding coordinate exchange must force its defining polynomial to vanish, i.e. $\sigma : y_i \mapsto -y_i $, so that $D_i=\{y_i=0\}$ is fixed. 
For point-wise fixed point with codimension larger than one, one needs to check whether the involution forces a subset of generators $\mathcal{F}\subseteq\mathcal{G}$ to vanish simultaneously. Namely, one needs to check $\mathcal{F}\cap\mathcal{G}_{-}\neq\emptyset$. It is important to note that the torus $\mathbb{C}^{*}$ actions provide $r=\text{rank}(\tilde{\mathbf{W}})$ additional degrees of freedom for the generators to avoid being forced to zero. In each subset of generators $\mathcal{F}$, we check for this by solving the system of equations
\bea
\lambda_{1}^{\tilde{W}_{1i}} \lambda_{2}^{\tilde{W}_{2i}} \dotsb \lambda_{r}^{\tilde{W}_{ri}} = \sigma(y_{i})/y_{i},\quad i=1,...,k' \, .
\eea
By the construction of the generator $y_{i}$, the right-hand side is equal to $\pm 1$. The set is point-wise fixed if this equation is solvable in the $\lambda_{i}$.

After finding out these naive fixed point loci, we need to check  whether each  point-wise fixed loci  lies in  Stanley-Reisner ideal $\cI_{SR}$.  The definition of  $\cI_{SR}$ leads $\cA$ to be split into {different patches \{$U_{i}$\}}.
 For a given fixed set, we compute in each sector $U_{i}$ the dimension of the ideal generated by the symmetry part of Calabi-Yau polynomial $P_{symm}$ and the fixed set generators $\mathcal{F}\equiv\{y_{1},...,y_{p}\}$
\bea
\label{eq:fixed}
\cI^{fixed}_{ip}=\langle U_{i},P_{symm},y_{1},...,y_{p}\rangle \, .
\eea
If  $\text{dim }\cI^{fixed}_{ip}<0$ for all $U_{i}$, then  $\cF$ does not intersect $X$. For each subset that is not discarded, we repeat this calculation for the ideal with one fixed set generator $\text{dim }\cI^{fixed}_{i1}$, and then two $\text{dim }\cI^{fixed}_{i2}$, etc. until { $\text{dim }\cI^{fixed}_{i\ell}=\text{dim }\cI^{fixed}_{ip}$} when adding more generators to the ideal no longer changes the dimension for any region $U_{i}$. Then, the intersection $\{y_{1}=\cdots =y_{\ell}=0\}$ of these generators gives the final point-wise fixed locus, with redundancies eliminated. In the end, an $O3$-plane corresponds to a codimension-3 point-wise fixed  subvariety, an $O5$- plane has codimension-2, and an $O7$-plane has codimension-1. If no O-planes exist and the invariant Calabi-Yau hypersurface is smooth, then the involution defines a { $\mathbb{Z}_{2}$ free action} on $X$. 

Finally, one can check whether the orientifold Calabi-Yau manifold support the string vacua, we consider a simple case where the $D7$-brane tadpole  cancellation condition is satisfied by  simply placing eight $D7$-branes on top of the $O7$-plane.   Then we only need to check the { $D3$-brane tadpole} condition which simplified to:
\bea
N_{D3} + \frac{N_{\text{flux}}}{2}+ N_{\rm gauge}= \frac{N_{O3}}{4}+\frac{\chi(D_{O7})}{4}\, \equiv - \, Q_{D3}^{loc}.
\eea
with $N_{\text{flux}}=\frac{1}{(2\pi)^4 \alpha^{'2}}\int H_3\wedge F_3 $, $N_{\text{gauge}}=-\sum_{a} \frac{1}{8\pi^2} \int_{D_a} \text{tr}{\cal F}_a^2$, and $N_{D3}$, $N_{O3}$ the number of $D3$-branes, $O3$-planes respectively.  
The $D3$-tadpole  cancellation condition requires the total D3-brane charge $Q_{D3}^{loc}$ of the seven-brane stacks and $O3$-planes to be an integer. If the involution passes this naive tadpole cancellation check, we will denote our geometry as a \lq\lq  {\it naive orientifold  Type IIB string vacuum}". One can further check the  smoothness of the orientifold Calabi-Yau and the Hodge number splitting under the involutions.

\section{Machine Learning for the Orientifold Calabi-Yau}
\label{sec:ML}
\subsection{Dataset and processing}
The database of orientifold Calabi-Yau threefolds ($h^{1,1}\leq 6$) we use to train our model was recently published in \cite{Altman:2021pyc} and we only explore the \lq\lq {\it geometry-wise proper involution}" which exchange the so-called proper non-trivial identical divisor (NID). Therefore, we will share the same assumptions as in \cite{Altman:2021pyc}, {\it i.e.} only consider favorable polytopes ($22974$ in total, with  $14$ admitting a non-trivial fundamental group). Among these $22960$ polytopes, there are $1401$ out of them  contain   {\it geometry-wise proper involution} we are interested in and $996$ out of  the $1401$ \lq\lq orientifolds" polytopes  admit  \lq\lq {\it  naive Type IIB string vacua}" (see \cite{Altman:2021pyc} for more details).

First of all, we use vertices of the favorable dual polytopes as the input data, which are matrices after putting together. There are two types of dual polytopes in the database: unresolved toric dual polytopes and resolved toric dual polytopes (which are more refined data). For the $h^{1,1}\leq 6$ case, we will use both data for comparison. Unfortunately, for $h^{1,1}> 6$ cases, there are no database for resolved polytopes yet, therefore we can only predict for these cases using the model learning from the unresolved dual polytopes \footnote{We will see in the next section, for $h^{1,1}\leq 6$   our results  from learning  both unresolved and resolved cases are almost identical, therefore we believe the predicted results for $h^{1,1}>6$ based purely on unresolved dual polytopes are reliable.}. The input data has different sizes as matrices, ranging from $4\times 5$ to $4\times 10$. To resolve this issue, we embed all the matrices into larger ones by adding zeros columns, and for the purpose of this paper (to make predictions for $h^{1,1}=7$ as example) we set the maximal size  as $4\times 11$. Below is one example how we embed a polytopy of Polyid$\# 2$ (in the database) into a  $4\times 11$ matrix by adding another six columns of zeros on the right:
\[
   \begin{bmatrix}
      -1 & -1 & -1 & -1 & 4  \\ 
      0 & 0 & 0 & 1 & -1  \\
      0 & 0 & 1 & 0 & -1  \\
      0 & 1 & 0 & 0 & -1  \\
   \end{bmatrix} 
   \: \overset{\text{embedding}}\longrightarrow \:
   \begin{bmatrix}
      -1 & -1 & -1 & -1 & 4 & 0 & 0 & 0 & 0 & 0 & 0 \\ 
      0 & 0 & 0 & 1 & -1 & 0 & 0 & 0 & 0 & 0 & 0 \\
      0 & 0 & 1 & 0 & -1 & 0 & 0 & 0 & 0 & 0 & 0 \\
      0 & 1 & 0 & 0 & -1 & 0 & 0 & 0 & 0 & 0 & 0 \\
   \end{bmatrix}      
\]

Second, the size of the dataset we can exploit is relatively small ($22960$) for machine learning and too small if we want to make predictions for $h^{1,1} \geq 7$ since the number of polytopes goes exponential as $h^{1,1}$ increases. One notable fact is that the ordering of the vertices in one data is only a matter of labeling and in principle we can always reorder these vertices. Therefore, we choose $120$ permutations to enlarge our dataset, which end up with   $2755200$ polytopes for machine learning. In practice, the permutations are realized by permuting columns of the original matrices.

The output data we are targeting would be: 1. Whether a polytope can result in  an orientifold  Calabi-Yau manifold with {\it geometry-wise proper involution}  and 2. Whether an \lq\lq orientifold" polytope from the first step can end up with a {\it naive (Type IIB) string vacua}. In either case, essentially this is a binary classification problem, valued in {\tt True} or {\tt False}. 

%In preparation for the machine learning in the next section, we split the data into three datasets: 1) training data, $50\%$, 2) validation data, $15\%$, and 3) testing data, $35\%$.

\subsection{CNN classifier and learning results}
\label{sec:ml}
\subsubsection*{Model building}
\label{sec:cnn}
One of the neural networks suitable for our classification problem is the convolutional neural network (CNN) (see for example \cite[Chapter~6]{Nielsen:2015nn} for an introduction). A typical CNN model consists of the input layer, convolutional layers, pooling layer(s), flatten layer, fully-connected layer(s) and the output layer. Due to our input data is of quite ``low resolution'' ($4\times 11$), we drop the pooling layer in our model. We add two fully-connected layers and each has $100$ neurons so that they can provide enough free parameters (weights and biases). But to prevent overfitting, we also add a dropout layer before the output layer. Since we are dealing with the classification problem, we choose the rectified linear activation function ({\tt ReLU}) as the activation function for most layers except for the output layer. 

We construct the above model using the well-established platform -- {\tt TensorFlow} \cite{Tensorflow}. In more specific, our model is defined as below:
\begin{itemize}
   \item Layers (excluded the input layer):
      \begin{itemize}
         \item[-] one $2D$ convolution layer, with $25$ filters, kernel size $3\times 3$ and {\tt ReLU} activation function,
         \item[-] one flatten layer, with default setup,
         \item[-] two full-connected layers (dense layers), both with $100$ neurons and {\tt ReLU} activation functions,
         \item[-] one dropout layer, with a dropout rate of $0.1$,
         \item[-] one output layer (dense layer), with $2$ neurons and {\tt Softmax} activation function.
      \end{itemize}   
   \item Loss function: {\tt Categorical Crossentropy}.
   \item Optimizer: {\tt Adam}, with default learning rate.
\end{itemize}
This model will be used to learn both orientifolds Calabi-Yau manifold and naive Type IIB string vacua. Note that being orientifold is an necessary condition for a space to be a naive Type IIB string vacuum, therefore we use the whole database ($h^{1,1}\leq 6$) to train the model to identify an \lq\lq orientifold" polytope  while use only orientifold data ($h^{1,1}\leq 6$ and {\tt orientifold==True}) to learn whether it can end up with the string vacua. Since we have enlarged the database by permutations, the data size of purely orientifolds is also considerably large enough ($1401\times 120 = 168120$) for machine learning. In practice, we can train the model for learning orientifold and string vacua separately as long as we use the restricted orientifolds database  to train the vacua classifier.

\subsubsection*{Results}
\label{sec:res}
\begin{figure}[!tbp]
   \centering
   \subfigure[Orientifold]{\includegraphics[width=0.45\textwidth]{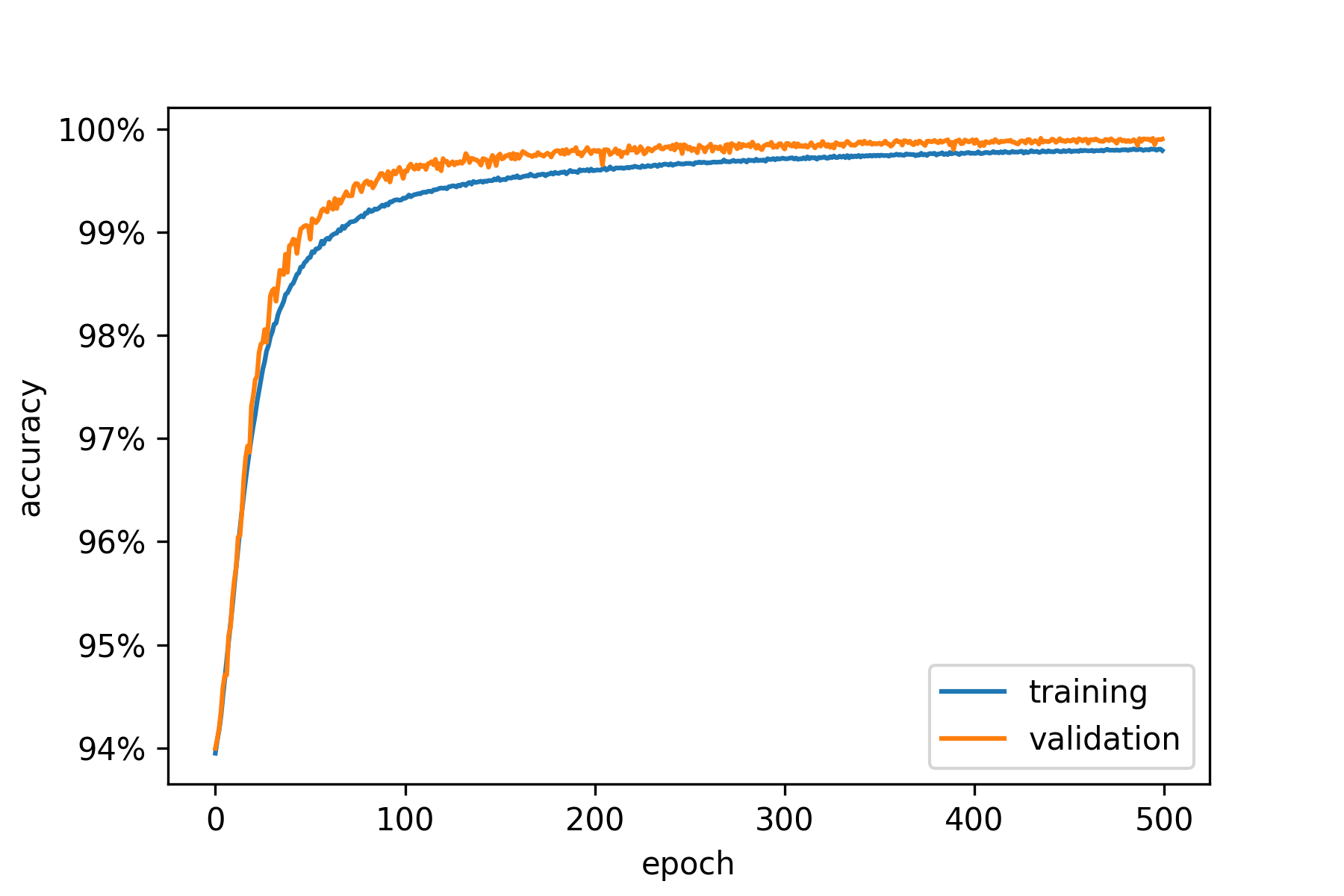}}
   \subfigure[Naive Type IIB string vacua]{\includegraphics[width=0.45\textwidth]{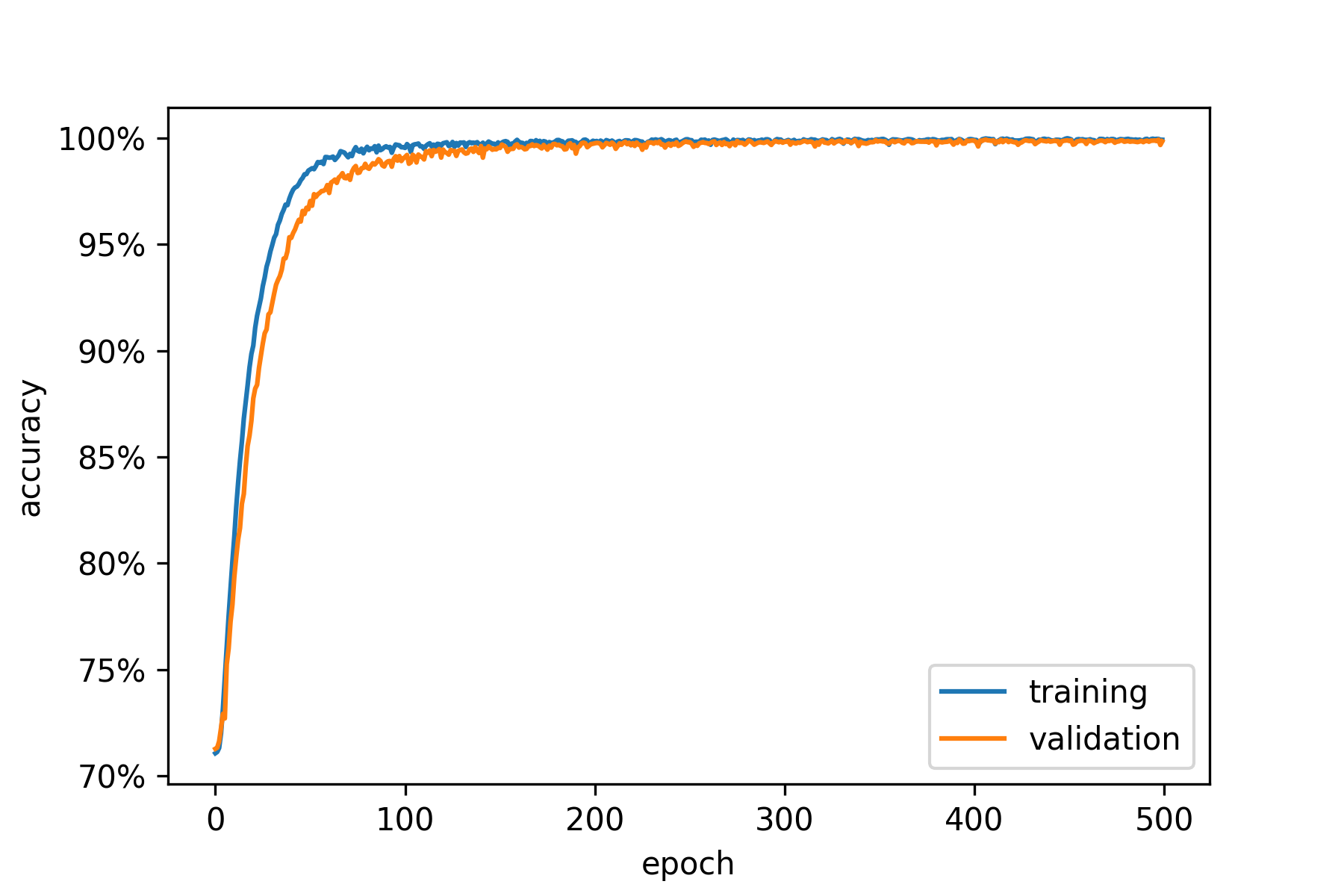}}
   \caption{Learning curves for unresolved data}
   \label{fig:learncurve}
\end{figure}

\begin{figure}[!tb]
   \centering
   \subfigure[Orientifold]{\includegraphics[width=0.45\textwidth]{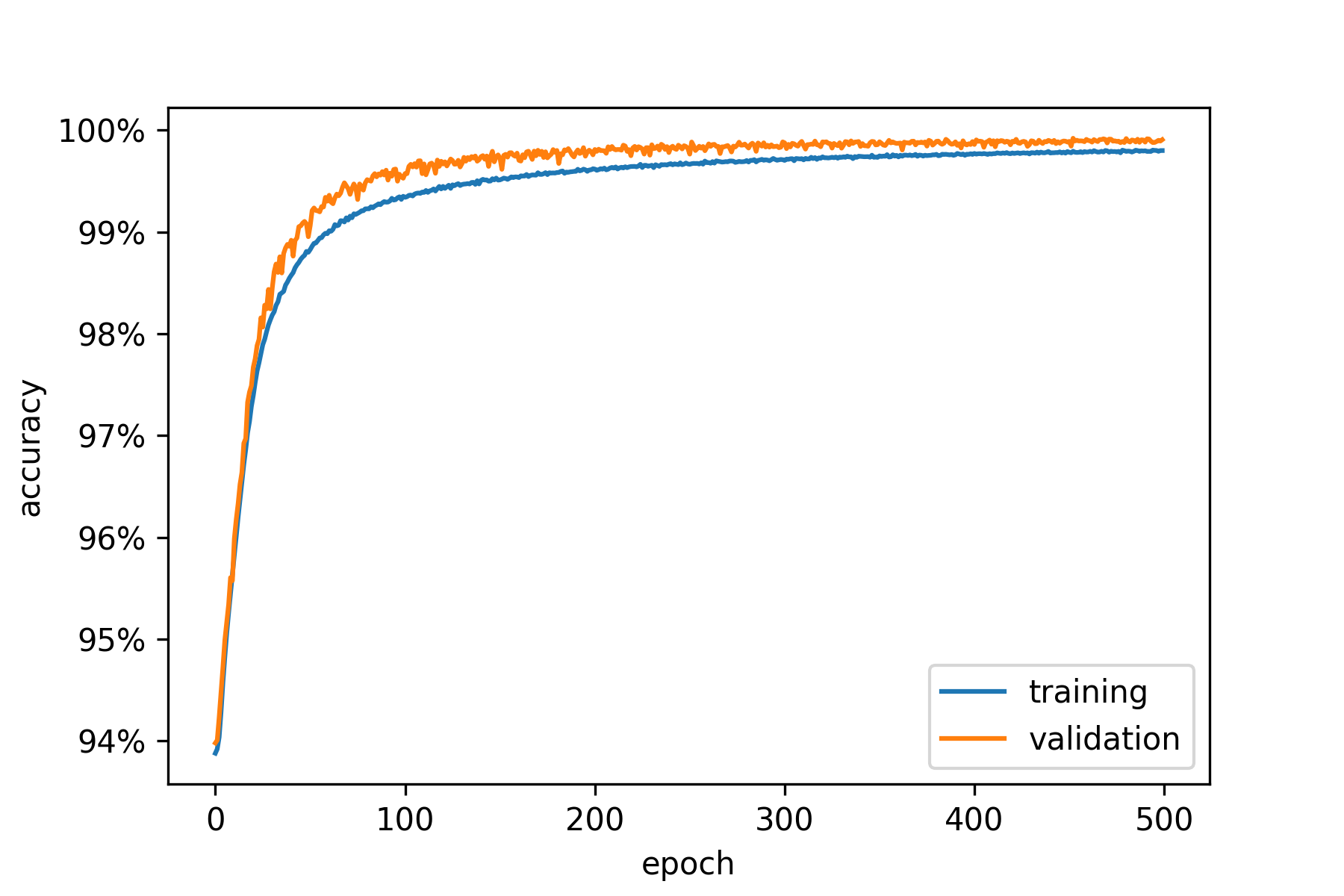}}
   \subfigure[Naive Type IIB string vacua]{\includegraphics[width=0.45\textwidth]{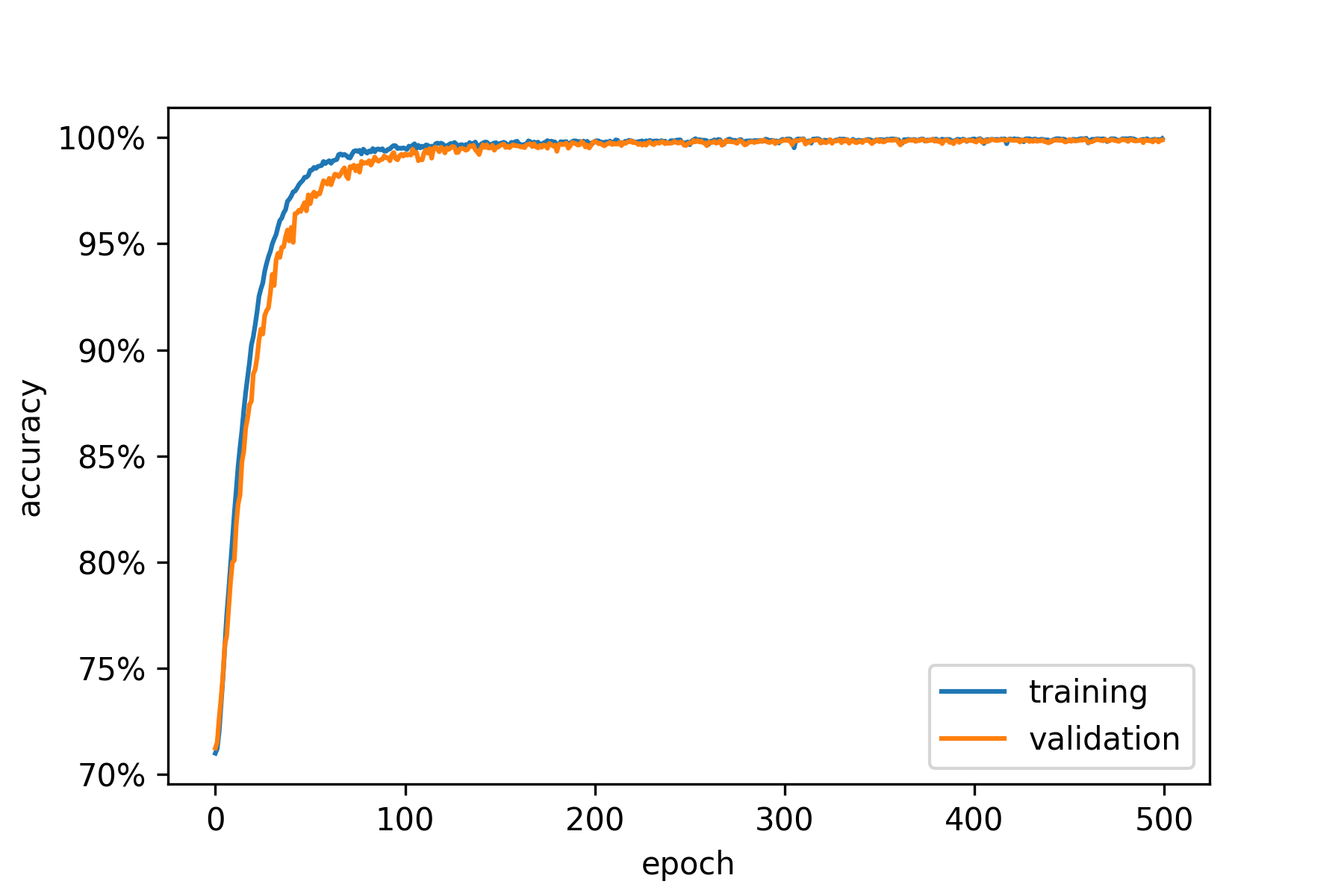}}
   \caption{Learning curves for resolved data}
   \label{fig:learncurveres}
\end{figure}

At this stage, we have feed the model with data of both resolved and unresolved toric dual polytopes. The test results are extremely accurate, $\gtrapprox 99.9\%$, as summarized in Table~{\ref{tab:result1}}, and they highly agree with the final training and validation accuracy (see Figure~\ref{fig:learncurve} for training unresolved vertexes and Figure~\ref{fig:learncurveres} for training resolved vertexes). The learning curves in Figure~\ref{fig:learncurve} also suggest that the high accuracy in our results is trustful and reliable, not obtained from overfitting. The learning curves for using resolved dual polytopes show the same features. 

\begin{table}[!h]
\centering
\renewcommand{\arraystretch}{1.5}
\begin{tabular}{ |P{6cm} |P{3cm}|P{3cm} | }
   \Xhline{1.5pt}
    & Unresolved  & Resolved\\\hline
   \hline 
   Orientifold & $99.906\%$ &$99.907\%$ \\\hline
   Naive Type IIB string vacua& $99.802\%$   & $99.897\%$ \\\hline
   \Xhline{1.5pt}
\end{tabular}\caption{Test results for $h^{1,1}\leq 6$.}
\label{tab:result1}
\end{table}

These test results also imply that, from the machine learning point of view, there would be almost no difference between using the resolved dataset and using the unresolved one when it classifies orientifolds or naive Type IIB string vacua. It is in this sense that we can confidently use  unresolved dataset to make predictions for higher $h^{1,1}\geq7$ in the next section.  We leave a full machine learning of resolved dual-polytopes in a future work.

\begin{figure}[!h]
   \centering
   \subfigure[Orientifold]{\includegraphics[width=0.45\textwidth]{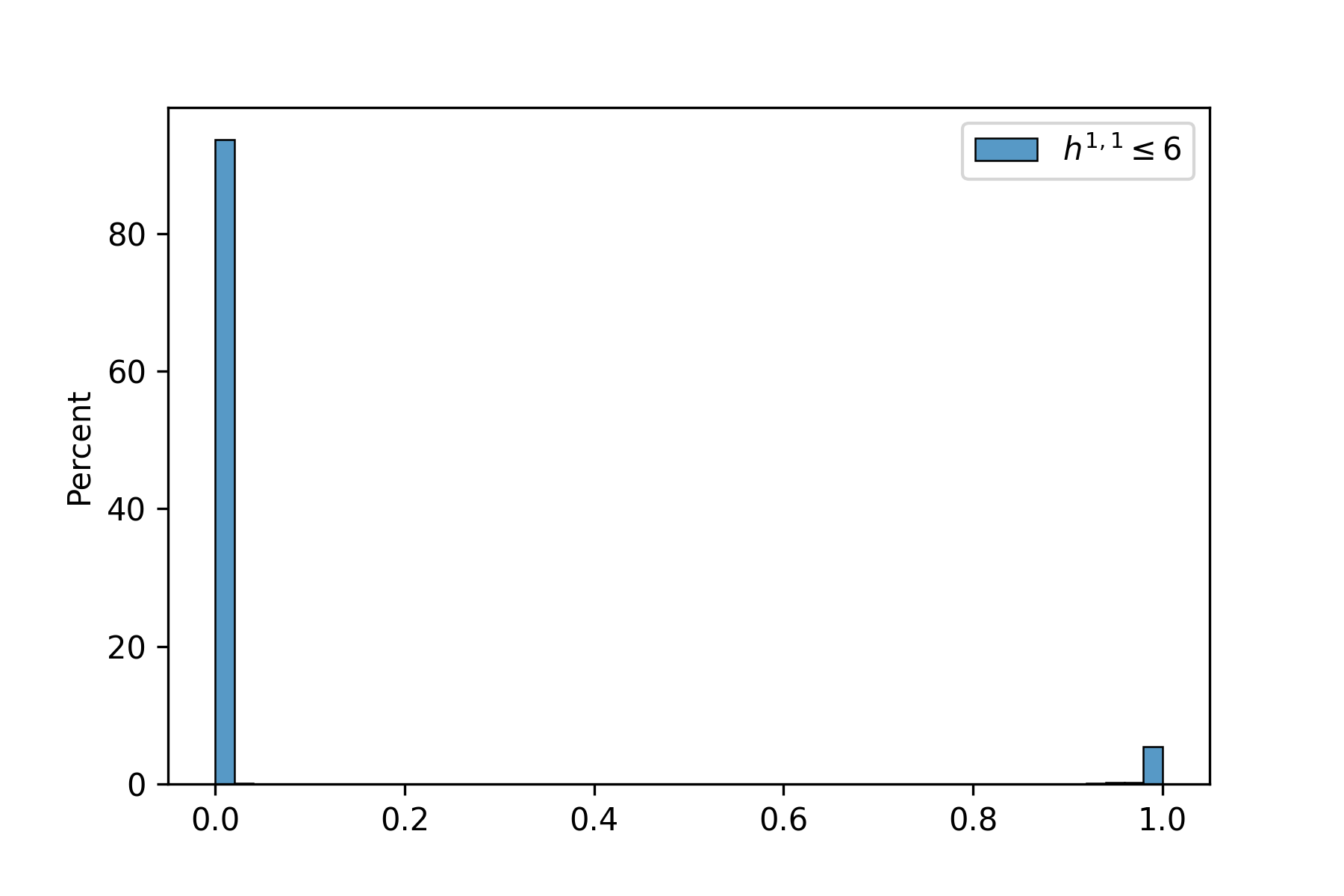}\label{fig:predictorint1}} 
   \subfigure[Orientifold (log-scale)]{\includegraphics[width=0.45\textwidth]{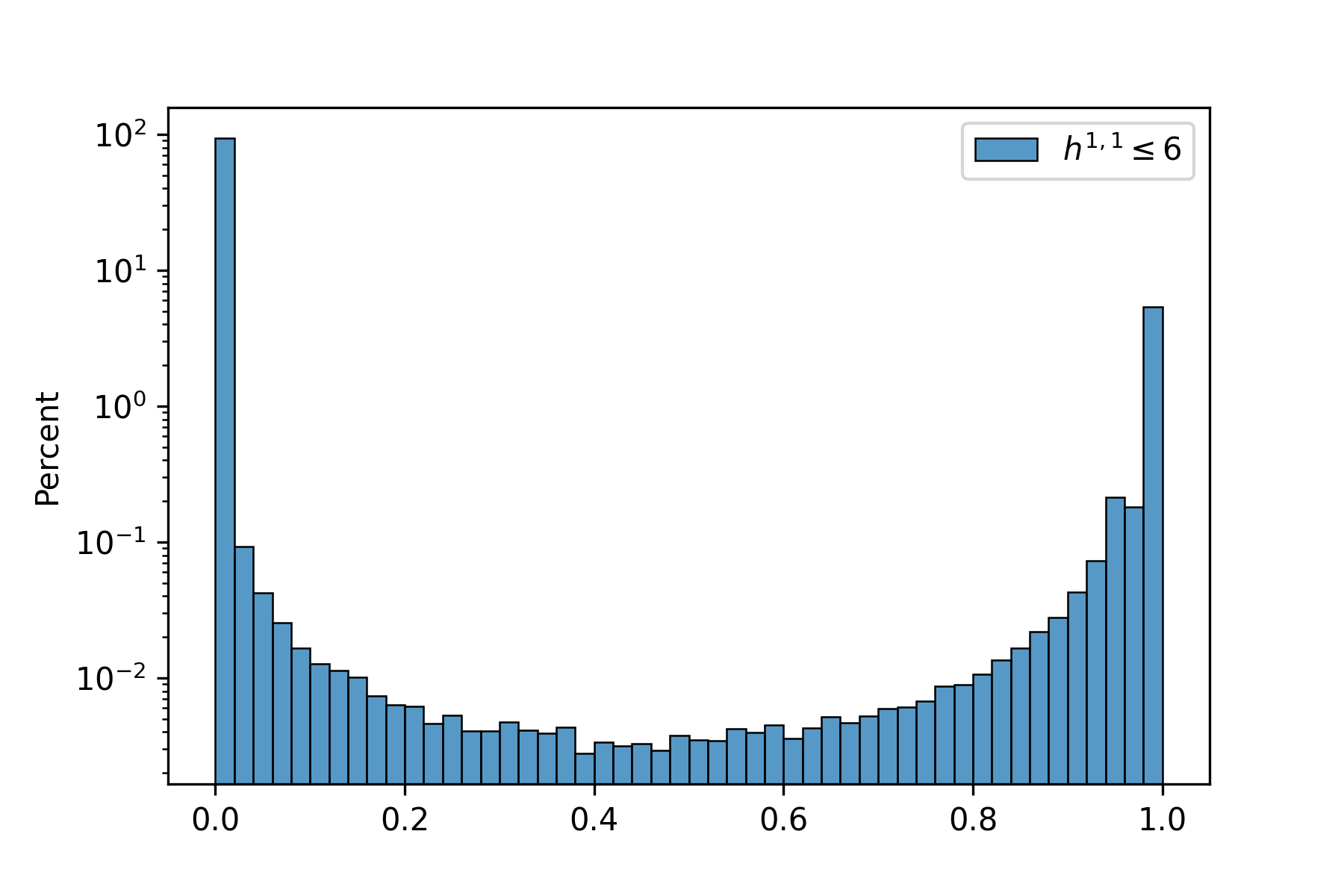}\label{fig:predictorint1log}} 
   \subfigure[Vacua]{\includegraphics[width=0.45\textwidth]{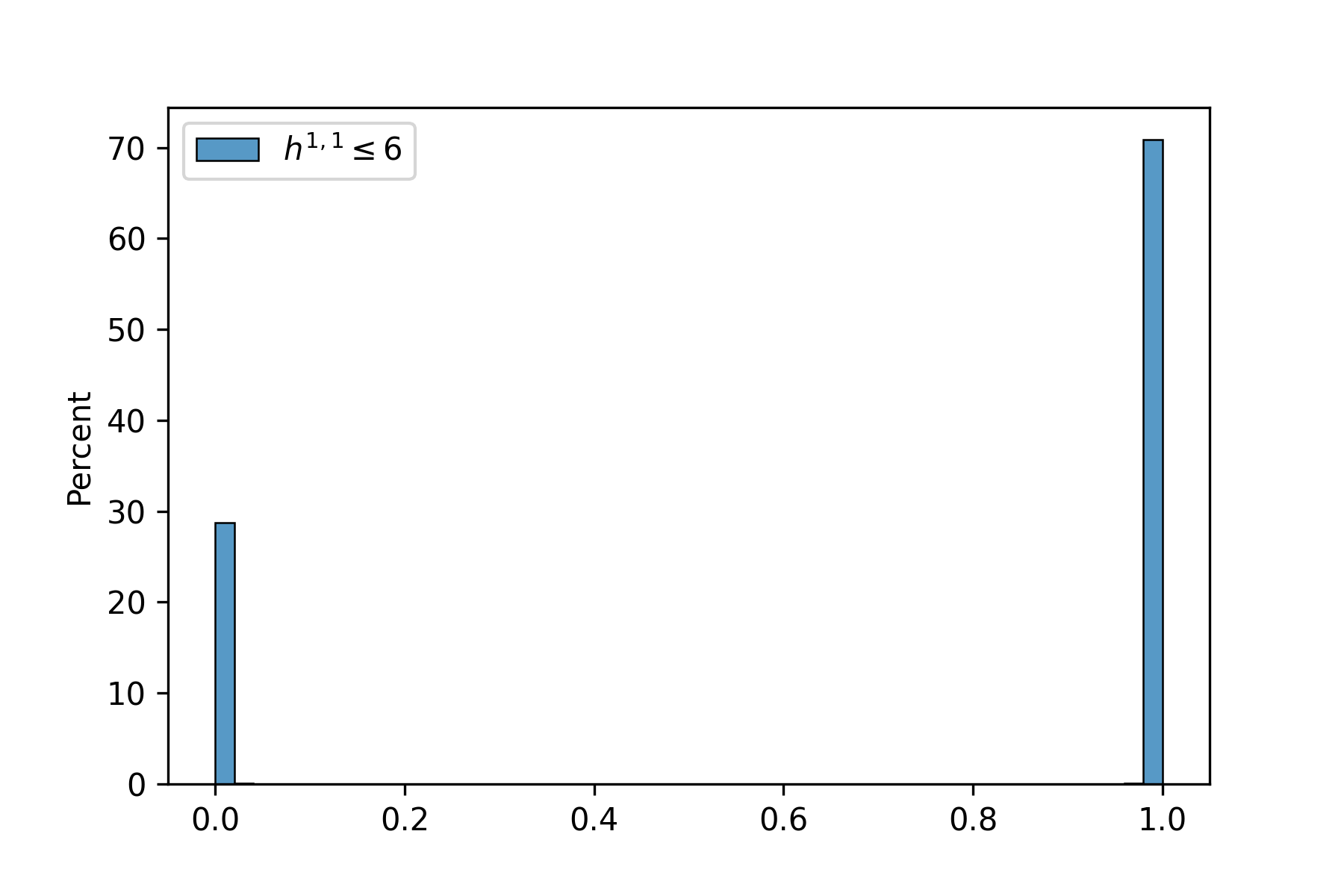}\label{fig:predictvacua1}}
   \subfigure[Vacua (log-scale)]{\includegraphics[width=0.45\textwidth]{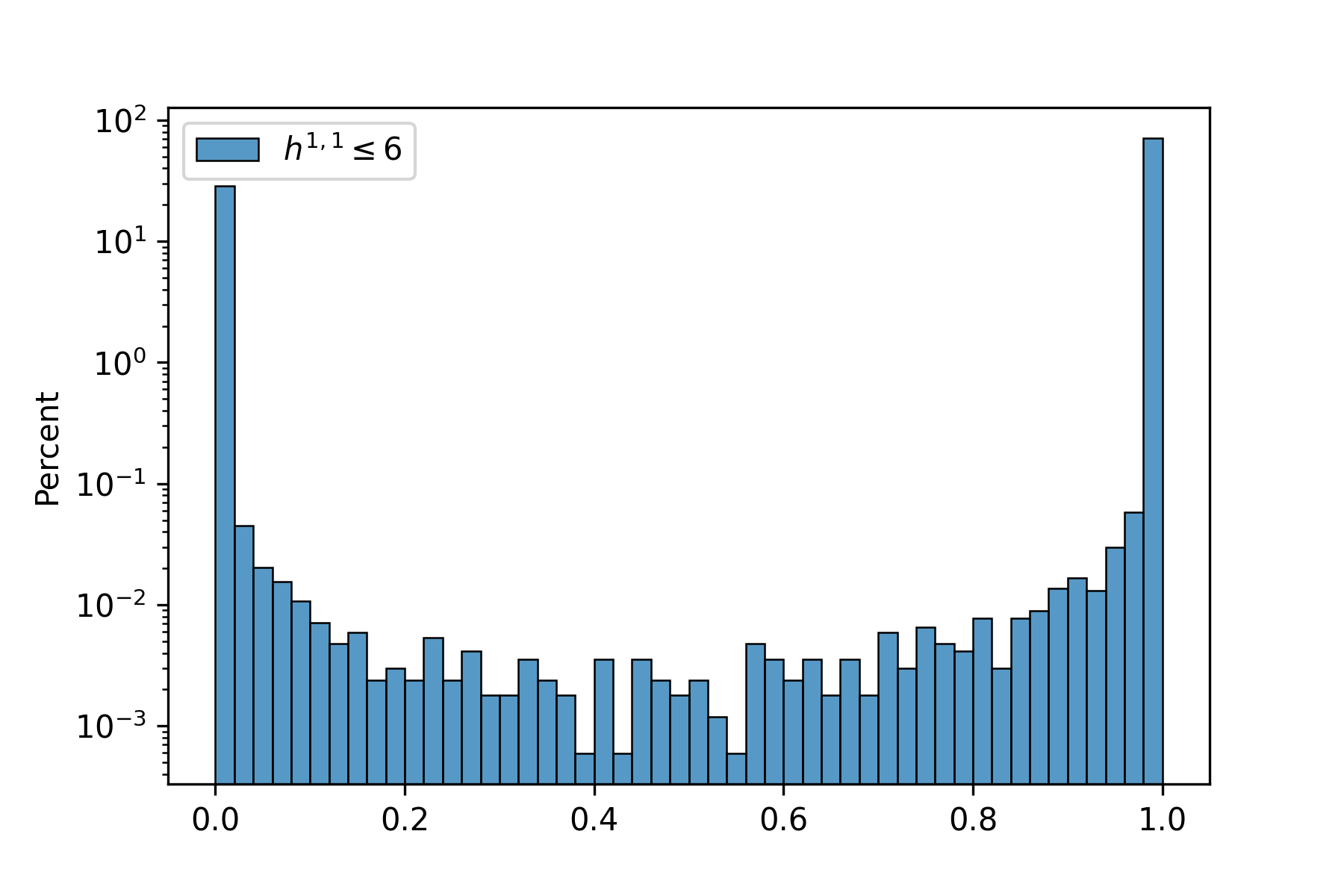}\label{fig:predictvacua1log}}
   \caption{Probability histograms for training data, obtained by evaluating the model with data of $h^{1,1}\leq 6$. $(a)$ and $(b)$ are orientifold distributions of the whole data set, while $(c)$ and $(d)$ are (naive Type IIB string) vacua distributions in all orientifolds (same meaning in Figure~\ref{fig:prediction}).}
   \label{fig:predictorint}
\end{figure}

The high accuracy in the learning results only can tell us that it is an accurate model. For an ideal binary classifying model, the output $\theta$ only take values in $\{0,1\}$ (corresponding to {\tt False} and {\tt True} respectively), namely there will be only two bins located in $0$ and $1$ separately in the distribution histograms. In practice, the output $\theta$ lies in the interval $[0,1]$ and the distribution histograms will have bins in between. We use the model to go through the training data again ($h^{1,1}\leq 6$) and see how our model recognize it. See Figure~\ref{fig:predictorint1} and Figure~\ref{fig:predictvacua1} and their corresponding distributions in log-scale. (For later comparison with dataset of different sizes, we draw the probability distribution histograms.) We should emphasis that all the orientifold distributions are drawn using the whole database, while all the (naive Type IIB string) vacua distributions are drawn only using orientifolds database.

In order to see whether it is a good binary classifier, in principle we should check its receiver operating characteristic (ROC) curve \cite{wiki:roc}. To put it simply, the farther the distributions of ``signal'' and ``background'' are separated, the better classifier it would be. In our scenario, the ``signal'' would indicate the orientifold or string vacuum. However, it is unrealistic to separate the signals from the background and what we really obtained is a combined signal-background distribution. In the probability distribution histograms, Figure~\ref{fig:predictorint}, there are two peaks, one is located at $0$ and the other one is located at $1$, while the bins in between are at least two orders of magnitude less (see the log-scale distributions). With the high accuracy from the learning results, we can tell that the peak located around $1$ is exactly where our ``signals'' are highly concentrated at. Meanwhile the ``background'' is highly concentrated around $0$. The histograms inform that the overlapping between the ``signal'' and ``background'' distributions is extremely small and therefore,  we are confident to claim that our neural network is an accurate and good classifier to pick out the polytopes which can result in an orientifold Calabi-Yau and string vacua.

This high accuracy indicates the orientifold symmetry, or more precisely the involution symmetry such as the Chow-group symmetry,  may already encoded in the polytope structure with unknown formula.    This is reasonable since at least for involution, we require it to be the symmetries of the graded Chow ring of the ambient space. It is very happy to see the machine learning seems to pick out this property quite efficiently. 

\section{Towards prediction for orientifold Calabi-Yau database with higher $h^{1,1}$}
\label{sec: high}

Another motivation of this paper is to search for potential  polytopes which may result in an orientifold Calabi-Yau, and further the string vacua with higher $h^{1,1}(X)$.  As we discussed in the Introduction of this paper, the difficulties including  too much amount of polytopes,  too many possible triangulations and involutions, complicated  and computationally intense algorithm, make us hard  to do a brute force calculation to scan the orientifold Calabi-Yau in the Kreuzer-Skarke database.   Due to the fact the orientifold signal is very rare in the full polytopes (around 5\% \cite{Altman:2021pyc}, see also Table.\ref{tab:stat}), it would be great even if we just train our machine to narrow down the candidate pool and  increase the successful rate by one order. 
Thus motivated by our learning result described in section \ref{sec:ml}, we  will try  to predict the possible polytopes with the desired property using our trained model. Here, our approach to achieve this goal is to utilize the classifier learnt from data with $h^{1,1} \leq 6$ to make predictions for $h^{1,1} \geq 7$.  
In this paper, we will only apply it to the $h^{1,1}=7$ case as an example.

\subsection{$h^{1,1} = 7$ case}

We evaluate the model using the (unresolved) dual polytopes with $h^{1,1} =7$, which can be obtained from the database \cite{Kreuzer:2000xy,database}. The number of these polytopes is $50376$ and it is much less than the size of data used to train our model ($ 50376 / 2755200 \sim 1.83\%$), and thus the parameters set by our training result is reliable to make predictions for $h^{1,1}=7$. 

\begin{figure}[!htb]
   \centering
   \subfigure[Orientifold]{\includegraphics[width=0.45\textwidth]{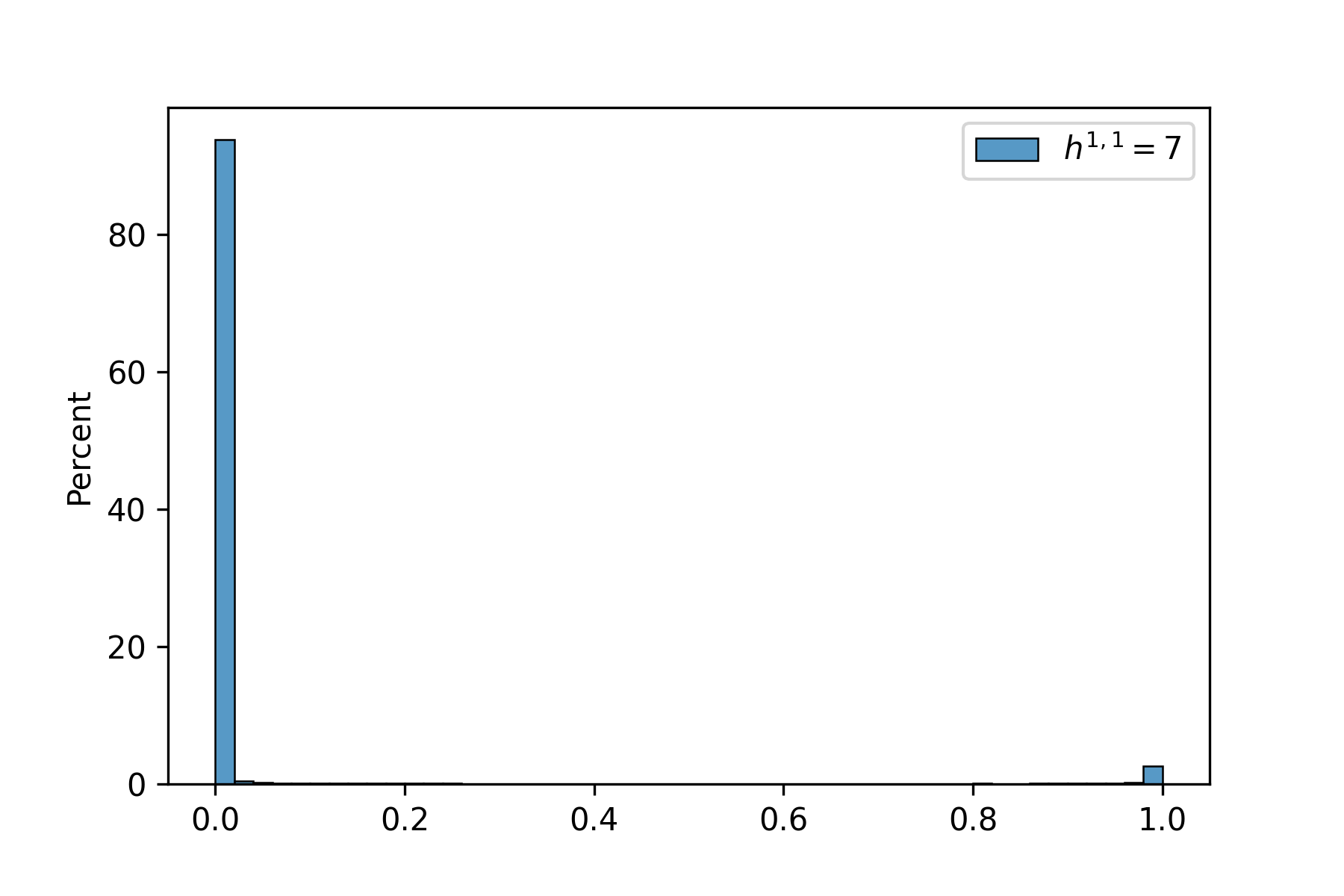}\label{fig:predictorint2}} 
   \subfigure[Orientifold (log-scale)]{\includegraphics[width=0.45\textwidth]{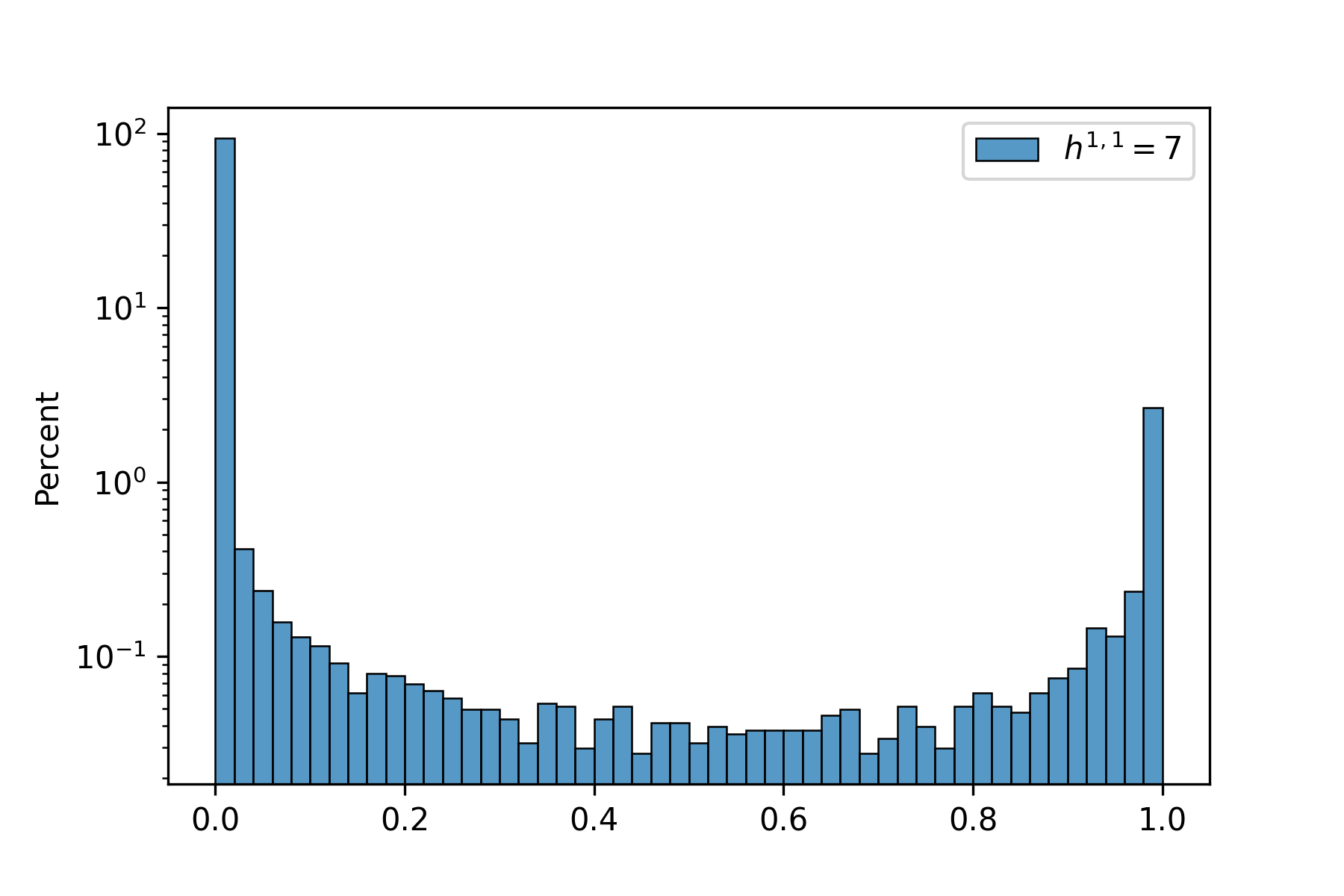}\label{fig:predictorint2log}} 
   \subfigure[Vacua]{\includegraphics[width=0.45\textwidth]{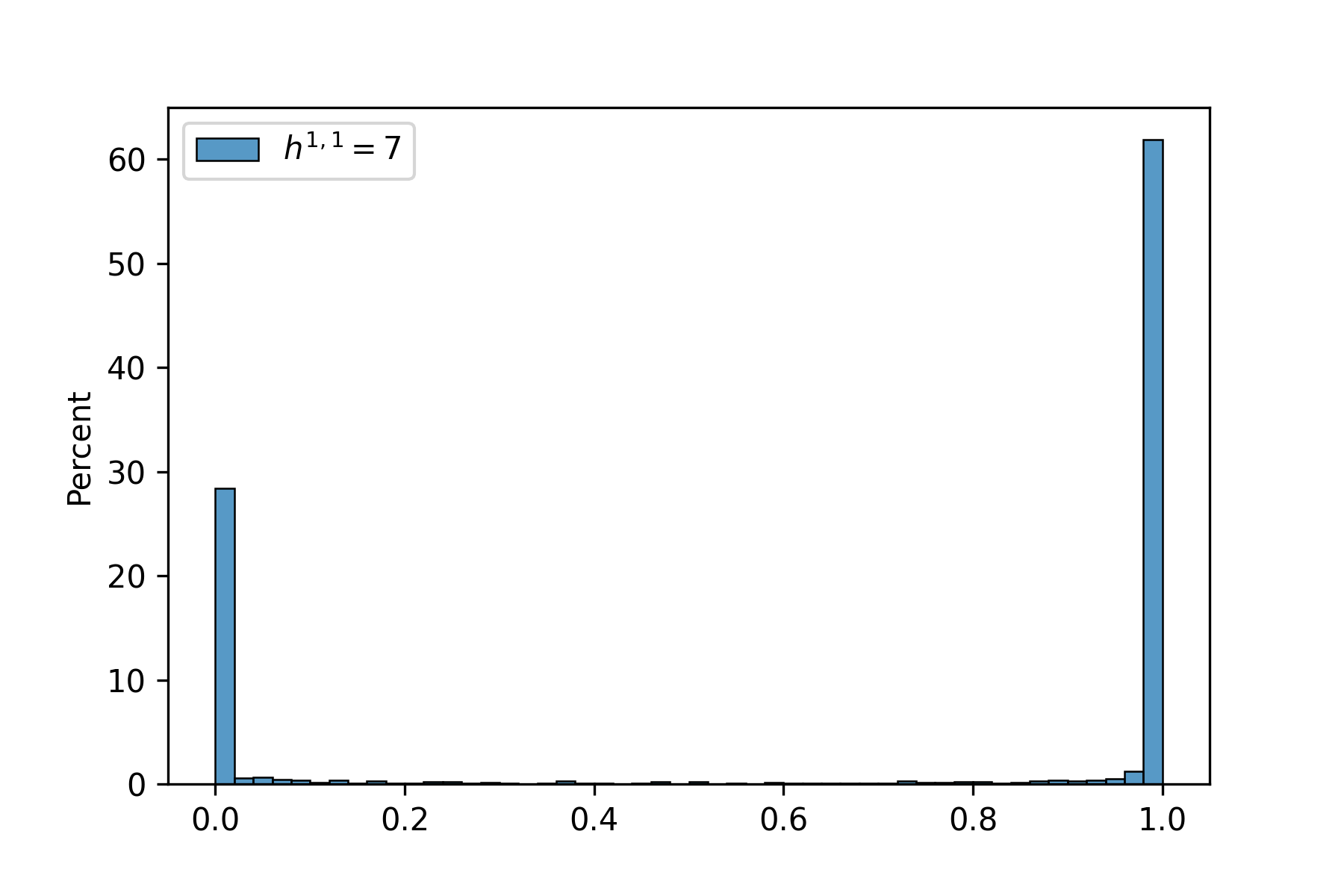}\label{fig:predictvacua2}}
   \subfigure[Vacua (log-scale)]{\includegraphics[width=0.45\textwidth]{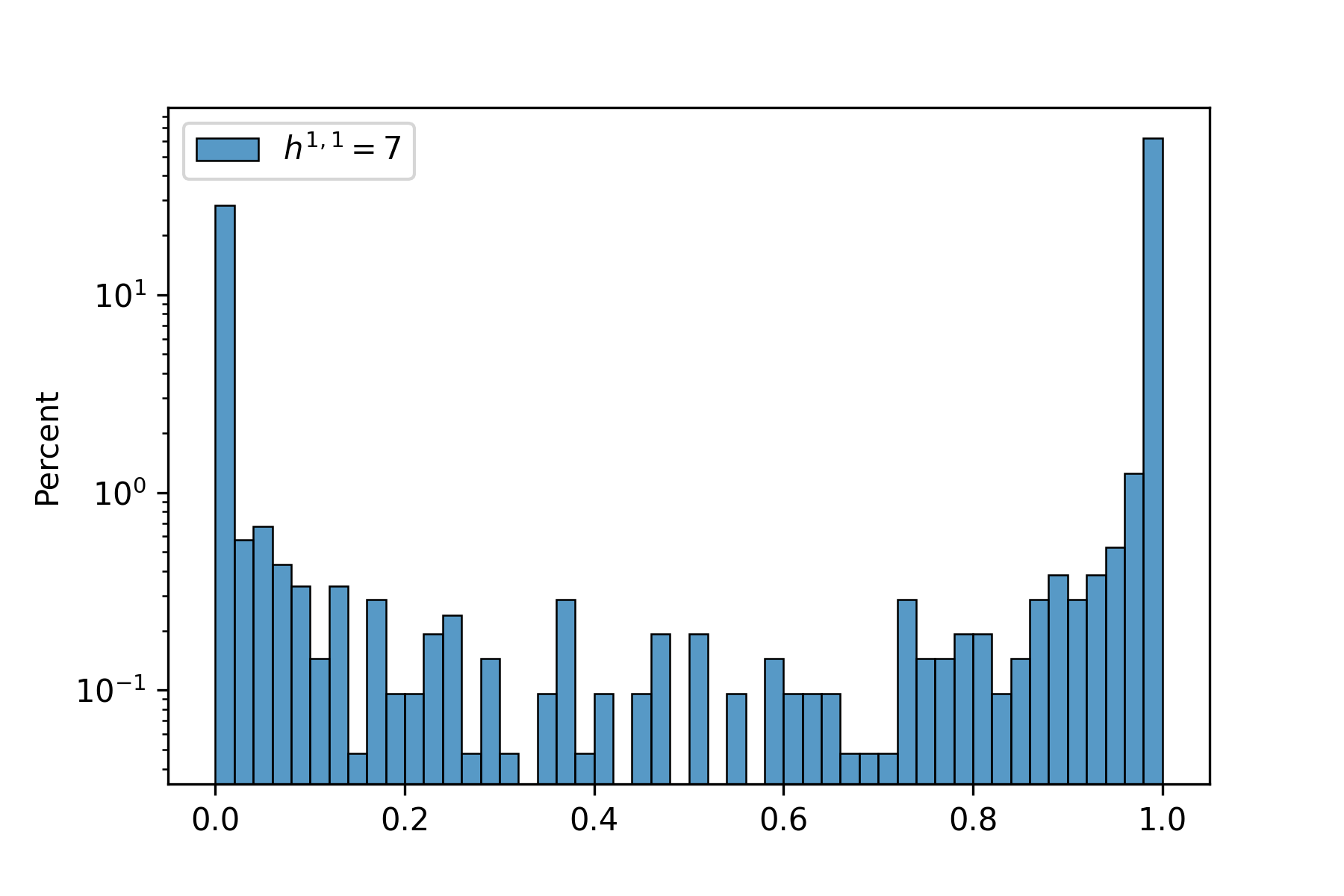}\label{fig:predictvacua2log}}
   \caption{Predicted probability histograms for data with $h^{1,1} = 7$.}
   \label{fig:prediction}
\end{figure}

The distributions of predictions are summarized in the probability histograms in Figure~\ref{fig:prediction}, and the shape of the corresponding probability histograms suggests it is still a very good classifier to  pick out the signal of candidates of \lq\lq orientifold" polytopes and string vacua. Compared with the histograms for $h^{1,1}\leq 6$, one can see that although those figures for $h^{1,1}=7$ are still with great shape, it is a little bit flattened. This is due to the following reason: we trained our model using favorable data but  directly applied the model on all data of $h^{1,1}=7$ without excluding the non-favorable ones since there is no such database available before a complicated calculation. 

 We choose the classifying threshold $\theta = 0.5$, which means the machine gives a orientifold (and further string vacua) whenever its output value is greater than $0.5$. With this choice,  the computer tells us that among the polytopes with $h^{1,1}=7$, there would be $2086$ of them may result in an orientifold Calabi-Yau manifold, $1399$ out of these  $2086$  polytopes may admit the naive Type IIB string vacua.  This is summarized in Table.\ref{tab:stat}. It is interested to notice that the percentage of polytopes which may contains the orientifold property indeed decrease following the trend when $h^{1,1}$ increase.

\begin{table}[h!]
  \centering
    \begin{tabular}{|r|r||P{0.7cm}|P{0.7cm}|P{1cm}|P{1cm}|P{1cm}|P{1cm}||P{1cm}|}
      \Xhline{1.5pt}
        \multicolumn{2}{|c||}{\parbox[c][2em][c]{6cm}{\centering$\mathbf{h^{1,1}(X)}$}} & \textbf{1} & \textbf{2} & \textbf{3} & \textbf{4} & \textbf{5} & \textbf{6}  & \bf{7} \\\hline\hline
            \multicolumn{2}{|r||}{\parbox[c][3em][c]{6cm}{\centering\textbf{\# of Trianed Polytopes}}} & 5     & 36    & 243   & 1185  & 4897  & 16608 & 50376\\   
    \hline
   \multicolumn{2}{|r||}{\parbox[c][3em][c]{6cm}{\centering\textbf{\# of \lq\lq orientifold"  Polytopes}}} & 0 & 1 & 16 & 96 & 330 & 958 & 2086 \\\hline
     \multicolumn{2}{|r||}{\parbox[c][3em][c]{6cm}{\centering\textbf{\% of \lq\lq orientifold"  Polytopes \\  }}} & 0 & 2.78 & 6.58 & 8.10 & 6.74 & 5.77 & 4.14 \\
   \Xhline{1.5pt}
       \end{tabular}
         \caption{Statistic counting on the polytopes which can result in orientifold Calabi-Yau. The result for $h^{1,1}\leq 6$ comes from \cite{Altman:2021pyc} while for $h^{1,1} = 7$ comes from our trained neural network.}
      \label{tab:stat}
    \end{table}

We have tested our predictions on a few examples and will present one specific example in the next section. These are limited examples among the data with $h^{1,1}=7$ that can actually be computed directly following the methods in \cite{Altman:2021pyc}. The reason is that we only considered favorable toric polytope in   \cite{Altman:2021pyc} while for $h^{1,1} \geq 7$ many of them are not.  Even the polytope is favorable, due to the large number of vertexes  in $h^{1,1}=7$, it is very hard to triangulate it to a smooth manifold in computing time.  Nevertheless, after the exact computations of some examples they do lie in our predictions of \lq\lq  orientifold" labeled by machine learning. We have attached a list of some favorable examples in Appendix~\ref{tab:predlist} which is classified as \lq\lq  orientifold".

\subsection{Predicted Example}

In this section, we present one particular example, which is labeled as both ``Orientifold'' and ``Vacua'' by machine learning.  Let us check in detail whether it gives the right answer.  The toric dual-polytope is given by the following $11$ vertices with Hodge number $h^{1,1} = 7, h^{2,1} = 53$.
\[
\begin{bmatrix}
   0 &1 &-1 &-1 &-1 &0 &0 &-1 &0 &-1 &1 \\
   0 &1 &-1 &0 &-1 &0 &0 &-1 &-1 &0 &0  \\
   0 &0 &-1 &-1 &-1 &-1 &0 &0 &0 &0 &1  \\
   -1 &1 &0 &-1 &-1 &0 &1 &-1 &0 &-1 &1\\
\end{bmatrix}.
\]

\noindent This example defines an MPCP desingularized ambient toric variety with weight matrix $\mathbf{W}$ given by
\bea
\label{eq:toricdata1}
\begin{array}{|c|c|c|c|c|c|c|c|c|c|c|}
\hline
x_{1} & x_{2} & x_{3} & x_{4} & x_{5} & x_{6} & x_{7} & x_{8} & x_9 & x_{10} & x_{11}\\\hline
1 & 1 & 1 & 0 & 0 & 0 & 0 & 0 & 0  & 1 & 1 \\\hline
1 & 0 & 0 & 0 & 0 & 0 & 1 & 0 & 0 & 0 & 0 \\\hline
0 & 1 & 0 & 0 & 1 & 0 & 0 & 0 & 0 & 1 & 1\\\hline
0 & 1 & 0 & 0 & 0 & 0 & 0 & 1 & 0 & 0 & 0 \\\hline
0 & 1 & 0 & 0 & 0 & 0 & 0 & 0 & 1 & 1 & 0\\\hline
0 & 0 & 0 & 1 & 0 & 0 & 0 & 0 & 0 & 0 & 1\\\hline
0 & 0 & 0 & 0 & 0 & 1 & 0 & 0 & 0 & 1 & 1\\\hline
\end{array}
\eea
\noindent There are $12$ different MPCP triangulations and we check the first one with Stanley-Reisner (SR) ideal as:
\bea
\label{eq:sr}
\cI_{SR}=\langle x_1 x_3,\, x_1  x_7,\, x_2 x_3, \, x_2 x_5,\, x_2 x_8,\,  x_2 x_9, \, x_4 x_8, \, x_4 x_9,\, x_4 x_{11},  \,x_6 x_8, \, x_6 x_{10}, \, x_9 x_{10},\, x_3 x_{11}, \, x_5 x_{11},\, \, x_5 x_7  \rangle. \nonumber
\eea
The Hodge numbers of the corresponding individual toric divisors $D_i \equiv \{x_i = 0\} $ can be calculate by  the
 \texttt{cohomCalg} package \cite{Blumenhagen:2010pv, cohomCalg:Implementation}: 
\bea
\label{eq:divhodge1}
 && h^{\bullet}(D_{1})= h^{\bullet}(D_{2}) = \{1,0,1, 20\}, \quad h^{\bullet}(D_{10})=h^{\bullet}(D_{11})=\{1,0,1,22\} \nonumber\\
&&  h^{\bullet}(D_{4})=h^{\bullet}(D_{8})=h^{\bullet}(D_{9}) =\{1,0,0,8\} \\
&&  h^{\bullet}(D_{3})=\{1,0,0,7\}, \quad h^{\bullet}(D_{5})=\{1,0,0,5\}, \quad h^{\bullet}(D_{7})=\{1,0,0,13\}  \nonumber
\eea
So there are several possible proper involutions. We first consider the involution as follows:
\bea
\label{eq:nontrivinvol}
\sigma: \,D_1 \leftrightarrow D_2, \,\, D_4 \leftrightarrow D_9,\,\,  D_{10} \leftrightarrow D_{11} \, .
\eea

In order to be a consistent orientifold, the volume form $\Omega_3$ should have a definite parity under $\sigma$. And, indeed we find $\sigma^* \Omega_3 = \, - \Omega_3$ and we would expect if there are any fix points under the involution, it should be $O3/O7$-planes. After a detailed calculation as described in section~\ref{sec:orientifold} and \cite{Altman:2021pyc}, we determined that there indeed are four $O7$-planes under the involution without any $O3$-plane.
\bea
 O7_{F_1}: x_4 x_{10} - x_9 x_{11},    \quad  O7_{F_2}: x_3,  \quad O7_{F_3}:  x_5,  \quad O7_{F_4}:  x_6. 
\eea

Then by placing eight $D7$-branes on top of the $O7$-plane, we only need to check the D3-tadpole cancellation condition:
\bea
N_{D3} + \frac{N_{\text{flux}}}{2}+ N_{\rm gauge}= \frac{N_{O3}}{4}+\frac{\chi(D_{O7})}{4} = \frac{36+9+7+12}{4} = 16
\eea
So we indeed get an Orientifold Calabi-Yau three folds and naive string vacua.

However, not all the involutions will end up with orientifold and string vacua.  For example,  if we consider other involutions like  $\{  D_4 \leftrightarrow D_9,  D_{10} \leftrightarrow D_{11} \}$,  we can get the fixed locus as three O7-plane  $\{ 
 \{  x_4 x_{10}- x_9 x_{11} \}, \{ x_6\},  \{x_5\} \}$ and one O3-plane located at $\{  x_3, x_7, x_4 x_{10}+x_9 x_{10}   \}$. But they can not satisfy the D3 tadpole cancellation condition.
 
 One should note that this example is a favorable one. To check whether one vertex is favorable or not, one should desingularize the vertex up to determinant singularity. For those toric Calabi-Yau which is not favorable, one could possible favor it following a similar method introduced in \cite{Anderson:2017aux} where all  Complete Intersection Calabi-Yau 3-folds (CICYs) \cite{Candelas:1987kf} can have a favorable description.   Then for favorable Calabi-Yau,  one should exhaust all triangulations to see whether there exist possible involutions and end up with fixed locus which may time consuming.  Then one can check whether the D3 tadpole  cancellation condition can be satisfied.
 
  One has to notice that the list presented in Appendix~\ref{tab:predlist}  are just predicted by the machine learning,  and one has to check whether these predictions is correct or not following the method described in this subsection. Usually this check is hard to be done and the precision is not very high even though the test accuracy is extremely high (for $h^{1,1} \leq 6$ it is around 99.9\%).  This is usually the case when one use some simple training data to predict a more complicated one,  like in the Kreuzer-Skarke database when  increasing the $h^{1,1}$.  We expect more than half of them will give the correct answer, i.e,  we successfully narrow down the candidate pool and  increase the successful rate by one order, from 5\% to 60\%.   This value is sensitive to the amount of training data if  it is not large. On the other hand, one may missing some orientifold polytope due to the classifying threshold $\theta$  one choose.  
    We have checked if we train our neural network for the database of $h^{1,1}\leq 5$ and do a prediction for $h^{1,1} = 6$,  it shows a similar picture. However, once we include more than  $10\%$ of the $h^{1,1} \leq 6$ data as training data, we would end up with a relatively very high accuracy of prediction (see Table.\ref{tab:result3}).  For  training the network with $10\%$ of the whole $h^{1,1} \leq 6$ database, the validation accuracy does not increase but fluctuate around $91\%$,  which indicates the initial setting of training ratio is too small. So one way to improve our prediction for $h^{1,1} = 7$ is to provide relatively small amount of  data ($> 10\%$)  for training.  Another way to improve the prediction is to perform a in principle unsupervised machine learning, such as  Generative Adversarial Network (GAN) \cite{Goodfellow} or Variational Autoencoder (VAE) \cite{Kingma}.   All of these difficulties need a more detailed and systemic study of the neural network, we will leave it for a further study.
   
\begin{table}[!h]
\centering
\renewcommand{\arraystretch}{1.5}
\begin{tabular}{ |P{6cm} |P{3cm}|P{3cm} |P{3cm} |  }
   \Xhline{1.5pt}
    Ratio of Training Data & $30\%$  & $20\%$ &  $10\%$  \\\hline
   \hline 
    Training Accuracy & $99.70\%$ &$99.64\%$  & $99.22\%$   \\\hline
     Validation Accuracy & $99.75\%$ &$99.16\%$  & $91.90\%$   \\\hline\hline
   Test Accuracy & $99.76\%$ &$99.14\%$  & $91.64\%$  \\\hline
   \Xhline{1.5pt}
\end{tabular}\caption{The test accuracy  varies according to the ratio of training data in $h^{1,1} \leq 6$.}
\label{tab:result3}
\end{table}

\section{Conclusion}
\label{sec:con}

In this paper, we use the machine learning  to clarify the polytope which can result in an orientifold Calabi-Yau hypersurface and together with the  ``{\it naive Type IIB string vacua}".  We show that indeed neural networks can be trained to give a high accuracy   (around  99.9\%)  for classifying the orientifold property and vacua.  This high accuracy indicates the orientifold symmetry, or more precisely the involution symmetry like the Chow-group structure,  may already encoded in the polytope structure with unknown formula.     In the end, we tried to use the trained neural networks model to go beyond the database and predict the orientifold property of polytope for higher $h^{1,1}(X)$.  Again, as being conservative, we should emphasize that some checks on the predictions still need to be done even though the machine learning has a extremely high accuracy on training data. 
In fact, our  work is  just a starting point for machine learning on the new orientifold Calabi-Yau database. There are several ways we can improve it and do the machine learning in a more systematic way.  

 First,  it would be very interesting to improve and generalize our work to  have a more systemic machine learning for higher $h^{1,1}$. One may try other neural networks like Generative Adversarial Network (GAN) or Variational Autoencoder (VAE) to improve the prediction.  Beside these in principle unsupervised training, we can still have a supervised training.   For example, to combine our method of finding orientifold  signal and the method of triangulation \cite{Long:2014fba} to generate enough training data for higher $h^{1,1}$  ($ \gtrsim 10\%$ of the target data) .   On the other hand, it would also be  interested to see whether learning the resolved dual-polytope vertex will give a more precise prediction in the higher $h^{1,1}$ case. For the machine learning for higher $h^{1,1}$ case, it would be great if the computer can pick out the polytopes which can result in a favorable Calabi-Yau and search for the orientifold structure there, since we only know the algorithm to do the brute force calculate the orientifold Calabi-Yau in the favorable case.

Second, in the context of CICY,  a landscape of orientifold vacua has been constructed~\cite{Carta:2020ohw, Carta:2021uwv} from the most favorable description of the CICY 3-folds database \cite{Anderson:2017aux}.  More general free quotients have been classified and studied in the case of CICYs~\cite{Braun:2010vc, Gray:2013mja, Candelas:2015amz, Constantin:2016xlj}.  Applying the machine learning technique to these geometry would also be great. In fact, a methodological study of machine learning on such kind of CICYs has been done in \cite{Erbin:2020tks}. In the coming paper\cite{cui2022}, we will show that such kind of  machine learning can also be done on  the so-called  \lq\lq generalized Complete Intersection Calabi-Yau" (gCICYs) \cite{Anderson:2015iia}.

Third, a lot of work has been made in understanding the statistical structure of the moduli in many classes of Calabi-Yau threefolds, without considering the orientifold involution explicitly,  such as the axion landscape or Swiss cheese structure \cite{Gray:2012jy, Long:2014fba, Galvez:2016qll, Long:2016jvd,  Altman:2017vzk,  Demirtas:2018akl, Halverson:2019cmy, Demirtas:2020dbm}.  Moreover, the study of the  landscape of Calabi-Yau manifold with $h^{1,1}_- \neq 0$ under the exchange involution is also very interesting \cite{Gao:2013rra, Cicoli:2021tzt, Carta:2021uwv}.  It would be great to combine our work to study the string model building in a real orientifold Calabi-Yau using machine learning technique. All of these works contain several technique problems and is important  and worthwhile for a further  study.

\section*{Acknowledgement}
The authors would like to thank    Wei Cui,  Jun Guo, Arthur Hebecker, Jinmian Li,  Andreas Schachner,  Juntao Wang for helpful discussions and correspondence.  XG was supported in part by  NSFC under grant numbers 12005150. 

\appendix
\section{ Some Predicted \lq\lq Orientifold" Polytopes  ($h^{1,1}=7$)}
\label{tab:predlist}
\begin{table}[H]
\renewcommand{\arraystretch}{1.2}
\renewcommand\normalsize{\footnotesize}%
\normalsize
   \centering
   \begin{tabular}{ P{14.2cm} |P{0.8cm}  }
\toprule
    Vertices &  Vacua \\
\midrule
$[[-1, -1, 0, 2], [-1, -1, 2, 1], [-1, -1, 2, 0], [-1, -1, 0, 1], [-1, 2, 0, 0], $\\$[1, -1, -1, 0], [-1, -1, 5, -1]]$ &\checkmark\\\hline
$[[0, 0, 0, -1], [1, 1, 0, 1], [-1, -1, -1, 0], [-1, 0, -1, -1], [-1, -1, -1, -1], [0, 0, -1, 0], $\\$[0, 0, 0, 1], [-1, -1, 0, -1], [0, -1, 0, 0],  [-1, 0, 0, -1], [1, 0, 1, 1]]$ &\checkmark\\\hline
$[[-1, 0, 0, -1], [-1, 0, -1, 0], [1, 1, 0, 0], [0, -1, 0, 0], [-1, -1, -1, -1], [-1, -1, 0, -1], $\\$[-1, -1, -1, 0], [0, 0, 0, 1], [0, 0, -1, 0],  [0, 0, 0, -1], [0, 0, 1, 0]]$ &\checkmark\\\hline
$[[-1, -1, 0, -1], [-1, -1, 0, 0], [-1, -1, -1, 0], [-1, 0, 0, 0], [-1, 0, 0, -1], [0, 0, 1, 0],$\\$ [-1, -1, -1, -1], [0, 0, 0, -1], [-1, 0, -1, -1], [0, -1, -1, 1], [1, 1, 0, 1]]$ &\checkmark\\\hline
$[[-1, 0, -1, 1], [-1, 0, -1, 0], [-1, -1, 0, -1], [-1, -1, -1, 0], [1, 0, 0, 0], [-1, -1, 0, 0], $\\$[-1, -1, -1, -1], [-1, 0, 0, -1], [0, -1, -1, -1], [0, 0, 0, -1], [1, 1, 1, 0]]$ & \\\hline
$[[-1, 1, -1, 1], [0, 1, -1, 1], [1, -1, 1, -1], [-1, -1, -1, 1], [-1, -1, 1, 0], [-1, -1, -1, 0], $\\$[0, 0, -1, 0], [1, 0, 1, -1]]$ &\checkmark\\\hline
$[[0, 1, 1, 1], [-1, -1, -1, -1], [-1, -1, -1, 0], [-1, -1, 0, -1], [0, -1, 0, -1], [-1, 0, -1, 1], $\\$ [-1, 0, 0, -1], [-1, 1, 1, 1], [0, 0, -1, 1], [1, 0, 0, -1]]$ &\checkmark\\\hline
$[[1, 1, 1, 1], [0, 0, -1, 1], [-1, -1, -1, -1], [-1, 0, 0, -1], [-1, -1, 0, -1], [-1, 0, -1, 1],$\\$  [-1, 1, 1, 1], [0, 0, 0, -1], [-1, -1, -1, 0], [0, -1, 0, -1]]$ &\checkmark\\\hline
$[[-1, 0, -1, 0], [1, 0, 0, 0], [0, -1, 0, -1], [-1, -1, 0, -1], [-1, -1, -1, -1], [0, 0, -1, 0], $\\$[1, 0, 0, -1], [0, 1, 0, 1], [-1, -1, -1, 0], [-1, 0, 0, 0], [0, 0, 1, 0]]$ &\checkmark\\\hline
$[[0, 0, -1, 0], [-1, -1, -1, 0], [-1, -1, -1, -1], [-1, -1, 0, -1], [-1, 0, -1, 0], [-1, 0, 0, -1], $\\$[-1, 0, 0, 0], [-1, -1, 0, 0], [0, -1, -1, 0], [0, -1, 0, -1], [1, 1, 1, 1]]$ & \\\hline
$[[-1, -1, -1, 0], [0, -1, 0, 0], [-1, 0, -1, 0], [0, 1, 1, 0], [-1, -1, 0, -1], [0, 0, 0, -1], [-1, 0, 0, -1],$\\$ [-1, -1, -1, -1], [1, 0, 0, 0], [-1, 0, 0, 0], [0, 0, -1, 1]]$ &\checkmark\\\hline
$[[0, 0, -1, 0], [-1, -1, -1, 0], [-1, -1, -1, -1], [-1, -1, 0, -1], [-1, 0, -1, 0], [-1, 0, 0, -1], $\\$[-1, 0, 0, 0], [-1, -1, 0, 0], [0, -1, -1, 0],[0, -1, 0, -1], [1, 1, 1, 1]]$ & \\\hline
$[[-1, -1, -1, 0], [0, -1, 0, 0], [-1, 0, -1, 0], [0, 1, 1, 0], [-1, -1, 0, -1], [0, 0, 0, -1], [-1, 0, 0, -1], $\\$ [-1, -1, -1, -1], [1, 0, 0, 0], [-1, 0, 0, 0], [0, 0, -1, 1]]$ &\checkmark\\\hline
$[[0, -1, 1, -1], [-1, 0, 0, -1], [0, -1, 0, -1], [0, 0, -1, 0], [-1, -1, -1, -1], [0, -1, -1, 0], [0, 1, 0, 1], $\\$[-1, -1, -1, 0], [-1, -1, 0, -1],[-1, 0, -1, 0], [1, 1, 0, 1]]$ &\checkmark\\\hline
$[[0, -1, 0, 0], [0, 0, -1, -1], [-1, 0, -1, -1], [-1, 0, -1, 0], [0, 0, 0, 1], [-1, -1, 0, 0], $\\$ [-1, -1, -1, -1], [-1, -1, -1, 0], [1, 1, 0, 0], [-1, -1, 0, -1], [0, 0, 1, 0]]$ &\checkmark\\\hline
$[[0, 0, -1, 0], [-1, 0, -1, -1], [-1, -1, 0, 0], [-1, -1, -1, 0], [1, 1, 0, 1], [0, -1, 0, 1], $\\$[0, 0, -1, -1], [-1, 0, 0, -1], [0, 0, 0, -1], [-1, -1, -1, -1], [0, -1, 1, 1]]$ &\checkmark\\\hline
$[[-1, -1, 0, -1], [0, 0, -1, 0], [-1, -1, -1, 0], [-1, -1, -1, -1], [0, -1, 0, 0], [0, -1, -1, 0], $\\$[-1, 0, 0, 0], [-1, 0, -1, -1], [0, 0, -1, -1], [1, 0, 1, 1], [0, 1, 1, 1]]$ &\checkmark\\\hline
$[[-1, 0, -1, -1], [0, -1, 0, 1], [-1, -1, 0, -1], [-1, 0, 0, -1], [-1, 0, 0, 0], [-1, -1, -1, -1], $\\$[0, 0, 0, -1], [-1, -1, 0, 0], [-1, -1, -1, 0], [0, -1, -1, 0], [1, 1, 1, 1]]$ & \\\hline
$[[-1, 0, -1, -1], [1, 1, 1, 0], [0, -1, 0, 0], [0, 0, 0, 1], [-1, 0, -1, 0], [-1, -1, -1, 0], $\\$[-1, -1, -1, -1], [-1, -1, 0, 0], [0, 0, 0, -1], [0, 0, -1, 0], [-1, -1, 0, -1]]$ &\checkmark\\
\bottomrule
\end{tabular}
\end{table}
\begin{table}[H]
\renewcommand{\arraystretch}{1.2}
\renewcommand\normalsize{\footnotesize}%
\normalsize
   \centering
   \begin{tabular}{ P{14.2cm} |P{0.8cm}  }
\toprule
    Vertices &  Vacua \\
\midrule
$[[-1, 0, -1, 2], [-1, -1, -1, 2], [0, 0, -1, 2], [-1, -1, -1, -1], $\\$[0, -1, 0, -1], [-1, -1, 0, -1], [-1, 0, 0, -1], [1, 1, 1, -1]]$ &\checkmark\\\hline
$[[-1, -1, 0, 2], [-1, -1, 1, 0], [0, -1, 0, 0], [-1, 2, -1, -1], [-1, -1, 0, 0],$\\$ [1, 2, -1, -1], [0, -1, 1, 0], [-1, 1, 0, -1], [0, 1, 0, -1]]$ & \\ \hline
$[[-1, 0, 0, -1], [-1, -1, -1, 1], [-1, -1, 0, -1], [-1, 0, -1, 0], [-1, 0, 0, 0],$\\$ [-1, -1, -1, -1], [0, -1, 0, -1], [0, 0, -1, 0], [1, 1, 1, 1]]$ &\checkmark\\\hline
$[[0, 0, 0, -1], [-1, -1, -1, 0], [-1, -1, 0, -1], [-1, -1, -1, -1], [-1, 0, 0, -1], [-1, 0, 0, 0], $\\$[0, -1, 0, 0], [-1, -1, 0, 0], [0, -1, -1, 0], [-1, 0, -1, 0], [1, 1, 1, 1]]$ & \\\hline
$[[0, -1, 0, -1], [-1, -1, -1, 0], [-1, 0, -1, 0], [-1, 0, 0, 0], [-1, -1, -1, -1], [-1, 0, 0, -1],$\\$ [0, -1, -1, 0], [-1, -1, 0, -1], [0, 0, -1, 0], [1, 0, 0, 1], [0, 1, 1, 0]]$ & \\\hline
$[[-1, 0, -1, 0], [0, 0, -1, 0], [-1, 0, 0, 0], [0, 0, 1, 0], [0, -1, 0, -1], [-1, -1, 0, -1], $\\$[0, -1, -1, 0], [-1, -1, -1, -1], [-1, -1, -1, 0], [-1, 0, 0, -1], [1, 1, 0, 1]]$ & \\\hline
$[[-1, 0, -1, -1], [-1, 0, 0, -1], [-1, 0, 0, 0], [-1, -1, -1, 0], [-1, -1, 0, -1], [-1, -1, -1, -1],$\\$ [0, -1, 0, -1], [-1, -1, 0, 0], [0, -1, -1, 0], [-1, 0, -1, 0], [1, 1, 1, 1]]$ & \\\hline
$[[-1, -1, -1, 0], [-1, -1, -1, -1], [-1, 0, 0, -1], [-1, -1, 0, -1], [-1, -1, 0, 0], [0, -1, -1, 1],$\\$ [-1, 0, -1, -1], [0, 0, 1, -1], [0, -1, -1, 0], [0, 1, 0, 0], [1, 0, 0, 1]]$ & \\\hline
$[[-1, -1, 0, -1], [0, 0, 0, 1], [0, -1, 0, 0], [-1, -1, -1, 0], [0, 0, -1, 0], [-1, 0, -1, -1], $\\$[-1, -1, -1, -1], [0, 0, -1, -1], [-1, -1, 0, 0],[0, 1, 0, 0], [1, 0, 1, 1]]$ &\checkmark\\\hline
$[[0, 0, -1, 0], [-1, 0, 0, -1], [-1, 0, -1, 0], [-1, -1, -1, 0], [0, -1, 0, 0], [-1, -1, -1, -1],$\\$  [-1, -1, 0, -1], [0, 0, 0, -1], [1, 1, 0, 1], [-1, 0, -1, 1], [0, -1, 1, -1]]$ &\checkmark\\\hline
$[[-1, 1, 1, -1], [-1, -1, 0, 0], [0, -1, -1, 1], [-1, 0, -1, 0], [-1, -1, -1, 1], [-1, -1, -1, 0],$\\$ [0, 0, -1, 0], [-1, 1, 1, 0], [0, -1, 0, 0], [1, 1, 1, 0]]$ & \\\hline
$[[0, 1, 0, 0], [-1, -1, -1, 0], [0, -1, 1, 0], [0, 0, -1, -1], [-1, -1, -1, -1], [-1, -1, 0, 0], $\\$[-1, 0, -1, -1], [-1, -1, 0, -1], [-1, 0, -1, 0], [-1, 0, 0, 0], [1, 1, 1, 1]]$ &\checkmark\\\hline
$[[0, 0, -1, 0], [-1, -1, 0, -1], [0, -1, 0, 0], [-1, -1, -1, -1], [-1, 0, -1, -1], [-1, -1, -1, 0],$\\$ [0, 0, -1, 1], [0, 1, -1, 0], [-1, 0, 0, -1], [1, 0, 1, 1], [0, -1, 1, -1]]$ &\checkmark\\\hline
$[[-1, 0, -1, 1], [-1, 0, -1, 0], [-1, 0, 0, -1], [-1, -1, 0, -1], [0, -1, 0, 0], [-1, -1, -1, 0], $\\$[-1, -1, 0, 0], [-1, -1, -1, -1], [0, 0, 0, -1], [0, -1, -1, -1], [1, 1, 1, 0]]$ & \\\hline
$[[0, 0, -1, -1], [0, -1, 0, 0], [0, 0, -1, 0], [0, 0, 0, 1], [-1, -1, -1, 0], [-1, -1, -1, -1], $\\$[-1, -1, 0, -1], [-1, 0, -1, -1], [1, 1, 0, 1], [-1, 0, 0, -1], [0, -1, 1, 0]]$ &\checkmark\\ \hline
$[[0, 0, -1, 0], [-1, -1, -1, 0], [-1, -1, 0, -1], [-1, 0, -1, 0], [-1, 1, 1, 1], [1, 1, 1, 1], [-1, 0, 0, 1],$\\$  [-1, -1, -1, -1],[-1, 0, 0, -1], [0, -1, 0, -1]]$ &\checkmark\\\hline
$[[-1, 0, 0, -1], [-1, -1, 0, -1], [-1, -1, -1, -1], [-1, -1, -1, 0], [0, -1, -1, 0], [0, 0, 1, 0], $\\$[-1, -1, 0, 0], [-1, 0, 0, 0], [0, -1, -1, -1], [-1, 0, -1, 0], [1, 1, 1, 1]]$ & \\\hline
$[[-1, 0, -1, -1], [0, 0, -1, 0], [-1, -1, -1, 0], [-1, 0, -1, 0], [0, -1, 0, 0], [-1, -1, -1, -1], $\\$ [-1, 0, 0, 0], [-1, -1, 0, 0], [-1, -1, 0, -1],[0, -1, -1, -1], [1, 1, 1, 1]]$ & \\\hline
$[[-1, -1, -1, 0], [-1, -1, 0, 0], [-1, 0, 0, 0], [0, -1, 0, 0], [-1, -1, 0, -1], [-1, 0, -1, 0], $\\$ [-1, -1, -1, -1], [0, -1, -1, -1], [0, 0, -1, 0], [-1, 0, 1, -1], [1, 1, 0, 1]]$ & \\\hline
$[[-1, 0, -1, 0], [-1, 0, -1, -1], [-1, -1, 0, -1], [-1, -1, 0, 0], [0, -1, 0, 1], [-1, -1, -1, 0], $\\$[-1, 0, 0, -1], [-1, -1, -1, -1], [0, 0, 0, -1], [0, -1, -1, 0], [1, 1, 1, 0]]$ & \\\hline
$[[-1, 0, -1, 1], [-1, 0, -1, 0], [-1, -1, 0, 0], [-1, -1, 0, -1], [-1, -1, -1, 0], [-1, 0, 0, -1],$\\$ [0, 0, 0, -1], [-1, -1, -1, -1], [0, -1, 0, -1],[0, -1, -1, 0], [1, 1, 1, 0]]$ & \\
\bottomrule
\end{tabular}
\end{table}
\begin{table}[H]
\renewcommand{\arraystretch}{1.2}
\renewcommand\normalsize{\footnotesize}%
\normalsize
   \centering
   \begin{tabular}{ P{14.2cm} |P{0.8cm}  }
\toprule
    Vertices &  Vacua \\
\midrule
$[[-1, 0, 0, -1], [-1, -1, -1, 0], [-1, -1, 0, 0], [0, -1, -1, 0], [-1, -1, 0, -1], [0, -1, 0, 0], $\\$ [0, -1, -1, -1], [-1, 0, 0, 0], [-1, -1, -1, -1], [-1, 0, -1, 0], [1, 1, 1, 1]]$ & \\\hline
$[[-1, -1, 0, 0], [-1, -1, -1, 0], [-1, 0, 0, -1], [-1, -1, -1, -1], [0, 0, 0, -1], [-1, -1, 0, -1], $\\$[0, -1, -1, -1], [-1, 0, 0, 0], [-1, 0, -1, 0], [0, 0, -1, 1], [1, 1, 1, 0]]$ & \\\hline
$[[-1, 0, -1, -1], [-1, 0, 0, -1], [-1, -1, 0, -1], [-1, -1, 0, 0], [0, -1, -1, 0], [-1, -1, -1, -1], $\\$[-1, -1, -1, 0], [-1, 0, 0, 0], [0, -1, -1, -1], [0, 1, 1, 0], [1, 0, 0, 1]]$ &\checkmark\\\hline
$[[-1, 0, -1, 1], [-1, -1, -1, 0], [-1, 0, 0, -1], [-1, -1, -1, -1], [-1, 1, 1, 1], [0, -1, 0, -1],$\\$ [1, 1, 1, 0], [-1, -1, 0, -1], [-1, 1, 1, 0], [0, 0, -1, 1]]$ &\checkmark\\\hline
$[[-1, -1, 1, 1], [-1, -1, 1, 0], [1, 1, 0, -1], [-1, 1, 0, -1], [-1, 1, -1, 0], [-1, 1, -1, -1], $\\$[0, -1, 0, 0], [0, -1, 0, 1], [-1, -1, 0, 0], [-1, -1, 0, 1]]$ &\checkmark\\\hline
$[[-1, 1, 0, -1], [0, -1, 1, 0], [0, -1, 0, 1], [-1, 1, -1, 0], [-1, -1, 0, 1], [0, 1, -1, -1], $\\$[-1, 1, -1, -1], [-1, -1, 0, 0], [-1, -1, 1, 0], [-1, -1, 1, 1], [1, 0, 0, 0]]$ &\checkmark\\\hline
$[[-1, 0, -1, 0], [-1, 0, -1, -1], [0, -1, -1, 1], [-1, -1, 0, 0], [-1, -1, 0, -1], [-1, 0, 0, -1],$\\$ [0, 1, 1, -1], [0, 0, 0, -1], [-1, -1, -1, -1],[-1, -1, -1, 0], [1, 0, 0, 1]]$ &\checkmark\\\hline
$[[0, -1, 1, 0], [0, -1, 0, 0], [0, -1, 0, 1], [-1, -1, 0, 0], [-1, -1, 0, 1], [-1, 1, -1, 0], $\\$ [-1, -1, 1, 0], [-1, 1, -1, -1], [-1, -1, 1, 1], [1, 0, 0, 0], [-1, 1, 0, -1]]$ &\checkmark\\\hline
$[[-1, -1, 0, -1], [-1, -1, -1, 0], [-1, 0, -1, -1], [-1, -1, -1, -1], [-1, 0, 0, 0], $\\$[1, 1, -1, -1], [1, 0, 1, 1], [0, -1, 1, 1], [-1, -1, 1, 1]]$ &\checkmark\\\hline
$[[-1, -1, 0, 3], [-1, -1, 0, 2], [-1, -1, 2, 0], [-1, -1, 1, 0], [1, -1, 0, 0], [-1, 2, -1, 0], [1, -1, 1, -1]]$ &\checkmark\\\hline
$[[-1, -1, 0, -1], [0, 0, -1, 0], [-1, -1, -1, 0], [-1, 0, -1, 0], [-1, 1, 0, 1], [-1, 0, 0, -1],$\\$ [0, -1, 0, -1], [1, 0, 0, 0], [-1, -1, -1, -1], [0, -1, -1, 0], [-1, 0, 1, 0]]$ &\checkmark\\
\bottomrule
\end{tabular}
\end{table}

\nocite{*}
\bibliography{ML}
\bibliographystyle{utphys}

\end{document}